%% file: mhdpre3.tex
%
%
%


\def\small{\scriptscriptstyle}
\def\sthirteen{\scriptscriptstyle 13}
\def\sfifteen{\scriptscriptstyle 15}
\def\stwelve{\scriptscriptstyle 12}
\def\snine{\scriptscriptstyle 9}
\def\sten{\scriptscriptstyle 10}
\def\sseven{\scriptscriptstyle 7}
\def\ssix{\scriptscriptstyle 6}
\def\stwo{\scriptscriptstyle 2}
\def\smalln{\scriptscriptstyle n}
\def\smallp{\scriptscriptstyle p}
\def\smalle{\scriptscriptstyle e}

\def\rhon{\rho_{\smalln}}
\def\rhop{\rho_{\smallp}}
\def\rhoe{\rho_{\smalle}}

\def\rhonn{\rho_{\smalln \smalln}}
\def\rhonp{\rho_{\smalln \smallp}}
\def\rhopn{\rho_{\smallp \smalln}}
\def\rhopp{\rho_{\smallp \smallp}}

\def\rpn{{\rhonp \over \rhon}}

\def\rpp{{\rhopp \over \rhop}}

\ifx\mnmacrosloaded\undefined \input mn\fi

%

\newif\ifAMStwofonts

\ifCUPmtplainloaded \else
  \NewTextAlphabet{textbfit} {cmbxti10} {}
  \NewTextAlphabet{textbfss} {cmssbx10} {}
  \NewMathAlphabet{mathbfit} {cmbxti10} {} 
  \NewMathAlphabet{mathbfss} {cmssbx10} {} 
  \ifAMStwofonts
    \NewSymbolFont{upmath} {eurm10}
    \NewSymbolFont{AMSa} {msam10}
    \NewMathSymbol{\upi}     {0}{upmath}{19}
    \NewMathSymbol{\umu}     {0}{upmath}{16}
    \NewMathSymbol{\upartial}{0}{upmath}{40}
    \NewMathSymbol{\leqslant}{3}{AMSa}{36}
    \NewMathSymbol{\geqslant}{3}{AMSa}{3E}

    \let\leq=\leqslant \let\le=\leqslant
    \let\geq=\geqslant \let\ge=\geqslant
  \else
    \def\umu{\mu}
    \def\upi{\pi}
    \def\upartial{\partial}
  \fi
\fi


\pageoffset{-2.5pc}{0pc}

\loadboldmathnames



\authorcomment{Preprint --  Accepted for publication in MNRAS}  

\begintopmatter  

\title{Magnetohydrodynamics in Superconducting-Superfluid Neutron Stars}
\author{Gregory Mendell}
\affiliation{Institute~for~Fundamental~Theory, University~of~Florida, Gainesville,~FL~32611, U.S.A.}
\affiliation{}
\affiliation{Email: mendell@phys.ufl.edu}

\shortauthor{G. Mendell}
\shorttitle{MHD in Superconducting Neutron Stars}


\abstract{MHD equations are presented for the mixture of superfluid neutrons,
superconducting protons, and normal electrons believed to exist in the
outer cores of neutron stars.  The dissipative effects of electron viscosity and
mutual friction due to electron-vortex scattering are also included. It is
shown that Alfv\'{e}n waves are replaced by cyclotron-vortex waves
that have not been previously derived from MHD theory.  The cyclotron-vortex
waves are analogous to Alfv\'{e}n waves with the tension due to the magnetic
energy density replaced by the vortex energy density.
The equations are then put into a simplified form
useful for studing the effect of the interior magnetic field on the dynamics.
Of particular interest is the crust-core coupling time which can be inferred
from pulsar glitch observations. The hypothesis that cyclotron-vortex waves
play a significant role in the core spin-up during a glitch is used to place
limits on the interior magnetic field.  The results are compared with those of
other studies.}

\keywords {dense matter -- hydrodynamics -- MHD -- stars: neutron -- pulsars: general.}

\maketitle  

\section{Introduction}

Below a superfluid transition temperature of approximately $10^9 \; {\rm K}$ 
the outer core of a neutron star is predicted to exist as an exotic plasma
of mostly superfluid neutrons, with a small fraction of type II superconducting
protons, normal electrons, and even fewer muons (Baym, Pethick
\& Pines 1969; Baym, Bethe \& Pethick 1971; Baym \& Pethick 1975, 1979; Epstein
1988).  The mass density of this region ranges from
$2.8 \times 10^{14}\; {\rm g} \cdot {\rm cm}^{-3}$
to a few times this value, and the protons and electrons make up
only a few percent of the total particle density.  Rotation causes the
superfluid neutrons to form an array of quantized vortices and
magnetic fields cause the superconducting protons to form an array
of quantized flux tubes (referred to generically as vortices from here on).
Pulsar observations show that neutron stars can rotate at least $642$ times per second
and typically have surface magnetic fields of about $10^{12}\; {\rm G}$ 
(Lyne 1995; Nice 1995).  Weaker fields in the range
$10^8$-$10^{10} \; {\rm G}$ have also been observed in neutron stars, mostly
in binary systems.  The evolution of the surface field and its link to the interior
field can provide clues to the physics of the neutron-star core (Bhattachary
1995).  Other clues come from thermal-evolution
studies (Tsuruta 1995) and the observation of rapid spin-ups, or glitches,
of several pulsars (Alpar \& Pines 1985;
Sauls 1989; Alpar 1995; Chau 1995; Link \& Epstein 1996.)

In the past it has been suggested that magnetic fields are responsible for the
core spin-up during a pulsar glitch and that glitches may
excite Alfv\'{e}n waves in this region (Baym, Pethick, Pines \& Ruderman 1969;
Van Horn 1980 and references therein). More recently, a constraint on the
crust-core coupling time for the Vela pulsar has been used to set lower
limits on the magnitude of the interior magnetic field (Abney, Epstein \& Olinto 1996).
These lower limits were derived using the Alfv\'{e}n wave velocity, though
a `modified' Alfv\'{e}n wave velocity was used in the case of a
superconducting-superfluid interior. 

The first goal of this paper is to develop the magnetohydrodynamic (MHD) 
theory of the superconducting-superfluid outer-core of a neutron star,
and to study the characteristic waves that occur in this exotic plasma.  
Hydrodynamic equations for this region
have been developed by Mendell \& Lindblom (1991), Mendell (1991a,b),
and Lindblom \& Mendell (1994). Here the MHD equations given in Mendell (1991a) are
developed with a focus on the effects of the proton vortices.
It is argued that the energy density due to individual proton vortices dominates
the vortex effects, and that the interactions between vortices,
including pinning between the neutron and proton vortices, can be ignored.  After the
MHD equations have been presented it is shown that Alfv\'{e}n waves do not occur
in the plasma considered here, but are replaced by cyclotron-vortex
waves that have not been previously derived from hydrodynamic theory.
The cyclotron-vortex waves found here are shown to be equivalent to
the `modified' Alfven waves used in Abney et al. (1996).

The second goal of this paper is to put the equations into a simplified form
useful for studying the effects of the interior magnetic field on the dynamics.
The main simplification is to drop all but the largest of the force terms due
to the magnetic flux and the vortices.  The resulting equations are the
superconducting-superfluid generalization of the MHD equations that
Easson \& Pethick (1979) gave for the case of an ordinary p-e plasma.
One advantage of the simplified equations is that the dissipative effects
of electron viscosity and mutual friction due to electron-vortex scattering are
easily studied following the methods given in Mendell (1991b).
The scattering of electrons off the neutron vortices usually dominates the dissipative
effects, though electron viscosity can be important in some situations.
Another advantage of the simplified equations is that they should be particularly convenient
for studying the effects of the magnetic field and cyclotron-vortex waves on the
core spin-up during a pulsar glitch. 

The final goal of this paper then is to investigate the hypothesis
that cyclotron-vortex waves play a significant role in the crust-core coupling time, which
can be inferred from pulsar glitch observations. Simple estimates are used to place
several limits on the interior magnetic field, and these are compared with the results
of other studies.

The paper is organized as follows.  In Section 2 the relevant properties
of superconductivity and superfluidity are reviewed and the MHD
equations for the outer-core region are presented.  The characteristic
frequencies associated with these equations are also reviewed. In Section 3 it
is shown that Alfv\'{e}n waves do not exist and the dispersion relation
for transverse axial modes in a constant magnetic field is found.
The frequency of these modes depends on the
cyclotron frequency and Kelvin vortex oscillation frequency, and so is called
the cyclotron-vortex frequency.  In Section 4 the effects of rotation are included
and it is shown that the usual inertial mode exists, along with an inertial mode
modified by Fermi liquid `drag' effects, and a mode dependent on the
cyclotron-vortex frequency.  In Section 5 the equations are simplified, and
dissipative effects are explored. In Section 6 the spin-up of the core during
a pulsar glitch is discussed. In Section 7 the results are briefly summarized
and suggestions for further work are discussed.

\section{Vortices and the MHD Equations}

An ordinary plasma is a mixture of oppositely charged fluids that is
locally charge neutral when in equilibrium.  High frequency
plasma waves result when the system is perturbed away from charge neutrality.
But low frequency waves are also possible and these waves can be studied by
taking the MHD limit.  This
limit is in force when the following restrictions apply:
1) the perturbations away from equilibrium preserve charge neutrality and,
2) the phase velocities are small compared to the speed
of light.  Alfv\'{e}n waves are the characteristic MHD waves that travel along
magnetic field lines in an ordinary plasma. (For more on the MHD limit
see Boyd \& Sanderson 1969; Jackson 1975; Easson \& Pethick 1979; Freidberg 1982.)

Now consider the plasma in the outer-core region of a rotating and magnetized
neutron star consisting of superconducting protons and normal electrons, immersed in
a background of superfluid neutrons. A major difference between this plasma
and an ordinary one is that the neutrons and protons are expected to form
arrays of quantized vortices (see especially Baym, Pethick \& Pines 1969 concerning
the proton vortices). Since the effects of the proton vortices play a dominate role in
the results to be discussed here, it will be useful to review their properties
before proceeding to the MHD equations. A more complete review is given
in Appendix A of Mendell (1991a).

In all the equations that follow
the subscripts $n$, $p$, and $e$ refer to the neutrons, protons, and electrons
respectively.  For each species, $\rho$ is the mass
density, $\bmath{v}$ is the velocity field, $m$ is the mass, and $e$, when
not a subscript, is the absolute value of the charge of the electron.
Standard symbols are used for the fundamental constants
$\hbar$, $c$, $k_{\rm {\small B}}$, and $G$, though $G$ is also
used as an abbreviation for Gauss, and Gaussian units are used throughout.

As in a laboratory type II superconductor (e.g., see Tilley \& Tilley 1986),
each proton vortex will carry one quantum of flux 
$$
\Phi_o = (hc/2 e) = 2.1 \times 10^{-7} \; {\rm G} \cdot {\rm cm}^2 , \eqno\stepeq
$$
and each vortex has a core region of normal protons with a radius
equal to the coherence length, given by
$$
\xi = {4 \hbar v_{{\rm {\small F}}p} \over 7 \pi
k_{\rm B} T_{{\rm c}p} } \approx 4.9 \times 10^{-12}
\left ( {m_p \over m_p^*} \right )
\rho_{p\sthirteen}{}^{1/3}
T_{{\rm c}p\snine}{}^{-1} \; {\rm cm} , \eqno\stepeq
$$
where $v_{{\rm {\small F}}p}$ is the Fermi velocity, $\rho_{p\sthirteen}$ is the mass
density of the protons divided by $10^{13} {\rm g} \; \cdot {\rm cm}^{-3}$, and
$T_{{\rm c}p\snine}$ is superfluid transition temperature divided by $10^9 \; {\rm K}$.
Outside the vortex core the proton velocity field and magnetic induction
fall off exponetially with a decay length-scale known as 
the London depth, given by
$$
\Lambda = \sqrt{{m_p^2c^2 \over 4 \pi e^2 \rhopp}}
\approx \, 9.3 \times 10^{-12}
\left ({\rhop \over \rhopp} \right )^{1/2}
\rho_{p\sthirteen}{}^{-1/2} 
\; {\rm cm} . \eqno\stepeq
$$
The assumption of type II superconductivity is valid
when $\sqrt{2} \Lambda > \xi$. In these equations $m_p^*$ is the effective
mass of the protons, while $\rhopp$ is a component of the mass
density matrix [see the discussion around eqs. (11) and (12)].
The energy per unit length of a proton vortex, $\varepsilon_p$,
is then specified by the kinetic energy of the protons plus the energy of
the magnetic induction, i.e.,
$$
\varepsilon_p = \int\int 
\left ( {1 \over 2} \rhopp v_p^2 + {1\over 8\pi}B^2
\right ) d a, \eqno\stepeq
$$
where $\bmath{v}_p$ is the velocity field of the protons, $\bmath{B}$ is the magnetic
induction of the vortex, and the integral is understood to extend over the plane
perpendicular to the axis of the vortex.  As is shown in standard texts on
superconductivity (and Mendell 1991a) $\varepsilon_p$ can be approximated as,
$$
\eqalign{
\varepsilon_p & \approx  {\Phi_o^2 \over 16 \pi^2 \Lambda^2}
\ln (\Lambda / \xi)  \cr
& \approx 2.1 \times 10^{6}
\left ( {\rhopp \over \rhop} \right )( 1 + \Delta)
\rho_{p\sthirteen}
\; {\rm erg} \cdot {\rm cm}^{-1} , \cr }
\eqno\stepeq
$$
where
$$
\Delta = 
1.6 \ln \left [ \left ( {m_p^* \over m_p} \right )
\left ( {\rhop \over \rhopp} \right )^{1/2}
T_{{\rm c}p\snine}
\rho_{p\sthirteen}{}^{-5/6} 
\right ] .  \eqno\stepeq
$$
This approximation is valid as long as $\Lambda > \xi$ which
implies $\Delta > -1$; it is assumed here that $\Delta$ is small or
at most of order unity. 

In a typical neutron star the spacing between
vortices is much larger than the London
depth but still much smaller than the radius of a neutron star. 
These statements apply to the neutron vortices as well as the
proton vortices.   Assuming a triangular array in equilibrium (as
is observed in terrestrial superfluids) the intervortex spacings
between neutron and proton vortices, $d_n$ and $d_p$, are given respectively by
$$
d_n = \Bigl( {h \over 2 \sqrt{3} m_n \Omega} \Bigr )^{1/2}
\approx 3.4 \times 10^{-3}
\Omega_{\stwo}{}^{-1/2} \; {\rm cm}, \eqno\stepeq
$$
$$
d_p = \Bigl( {h \over  \sqrt{3} m_p 
\Omega_{\rm {\small C}} } \Bigr )^{1/2}
\approx 3.5 \times 10^{-10}
B_{o\stwelve}{}^{-1/2} \; {\rm cm}, \eqno\stepeq
$$
where $\Omega_{\stwo}$ is the rotational angular velocity divided by
$10^2 \; {\rm s}^{-1}$, $B_{o\stwelve}$ is the average equilibrium magnetic
induction divided by $10^{12} \; {\rm G}$,
and $\Omega_{{\rm {\small C}}}$ is the cyclotron frequency of the protons.

To investigate the hydrodynamic behavior of the fluids
on length-scales much larger than the intervortex spacing it is therefore
appropriate to perform a smooth averaging over the vortices, 
as discussed e.g., by Baym \& Chandler (1983) (for a review see Sonin 1987).
Following these references the number of neutron and
proton vortices per unit area, $n_{{\rm v}n}$ and $n_{{\rm v}p}$,
are related to the smooth-averaged vorticities and magnetic induction
by the following form of the London equations:
$$
{\bomega}_n = {h \over 2m_n}n_{{\rm v}n}{\bnu}_n, \eqno\stepeq
$$
$$
{\bomega}_p + a_p\bmath{B} = {h \over 2m_p}n_{{\rm v}p}{\bnu}_p.
\eqno\stepeq
$$
In these equations $a_p = e/m_p c$, and the unit vectors ${\bnu}_n$ and
${\bnu}_p$ point in the smooth-averaged direction of the vortices.
In equation (10) $\bmath{B}$ is the smooth-averaged magnetic induction.
The smooth-averaged vorticities in these equations are given by 
${\bomega}_n = {\bnabla} \times \bmath{v}_n$ and
${\bomega}_p = {\bnabla} \times \bmath{v}_p$, where $\bmath{v}_n$ and
$\bmath{v}_p$ are the smooth-averaged superfluid velocities.  The London
equations then can be taken as the definition of these velocities.  Given
these definitions it must be noted that due
to Fermi liquid `drag' effects the mass currents of the neutrons
and protons depend on the superfluid velocities of both
species via the mass density matrix $\rhonn$, $\rhonp = \rhopn$, and $\rhopp$,
as discussed by Andreev \& Bashkin (1976) and Alpar, Langer \& Sauls (1984).
The recent calculation of Borumand, Joynt \& Klu\'{z}niak (1996) shows that
$$
\rhonp = \rhopn = \rhop \left [ 1 - 
{m_p \over m_p^*} (1 + F_1^{pp} / 3) \right ], \eqno\stepeq
$$
$$
\rhonn = \rhon - \rhonp,
\qquad 
\rhopp = \rhop - \rhopn, \eqno\stepeq
$$
where $m_p^*$ and $F_1^{pp}$ are the effective mass of the proton
and a generalized version of the Landau parameter respectively.  These factors are
defined in Fermi liquid theory in terms of quasiparticle interactions that must
be determined either from nuclear matter theory or by observation.  Most of the previous
studies of superfluidity in neutron stars have ignored the factor $F_1^{pp}$.  However,
because of the uncertainties in $m_p^*$ and $F_1^{pp}$, all studies agree that for the
neutron star outer core $\rhonp \approx - \rhop$, $\rhonn \approx \rhon$,
and $\rhopp \approx 2 \rhop$.
The main results given in this paper do not depend on the occurence of the `drag' effect,
though components of the mass density matrix will appear in many of the equations. (The
exception being that the `drag' effect greatly increasing the scattering of electrons
off the neutron vortices.)

Throughout the rest of this paper all quantities
will be understood to represent smooth averages over length scales large
compared to the inter-vortex spacing but small compared to the
length scales of physical interest (e.g., the stellar radius).
In the smooth-averaged theory the effects of the underlying vortices
manifest themselves as vortex forces acting on the smooth-averaged
flow of the superfluids. Bekarevich \& Khalatnikov (1961) show
a form for these forces can be found by allowing
the energy density of the fluid to depend on the vortex densities.
Mendell \& Lindblom (1991) extended their
method to mixtures of charged superfluids including the
electromagnetic coupling. Following these references, first the
vector fields ${\blambda}$ are defined so that the energy density due
to the vortices, $U_{{\rm {\small V}}}$, has a differential
$$
{\bf d} U_{{\rm {\small V}}} 
=  {h \over 2m} {\blambda} \cdot {\bf d} \, n_{{\rm v}} \, {\bnu}. \eqno\stepeq
$$
It can then be shown that a vortex force per unit mass given by
$$
\bmath{f}_{{\rm {\small V}}} = {1 \over \rho}({\bnabla} 
\times {\blambda}) \times  {h \over 2m} n_{{\rm v}}{\bnu} \eqno\stepeq
$$
acts on the smooth-averaged flow of the superfluid. 

It should be noted that the Bekarevich \& Khalatnikov (1961) model ignores
interactions between the vortices.  Since the magnetic induction and circulation
of a proton vortex falls off exponentially, ignoring the interactions between the
proton vortices should be an acceptable approximation as long as the separation
between these vortices is much larger than the London depth i.e., $d_p >> \Lambda$.
For simplicity interactions between neutron vortices will also be
ignored here as including them would not significantly effect the results
of this paper.\note{${}^\star$}{Interactions between the neutron vortices
can give rise to low-frequency oscillations of the vortex array known as
Tkachenko waves (Tkachenko 1966; Ruderman 1970; Fetter \& Stauffer 1970; Sonin 1987).}
Finally, interactions between proton and neutron vortices, which can lead to pinning
(or interpinning) of these vortices, must be considered.  First note
that Alpar, et al. (1984) have shown that the `drag' effect causes the
neutron vortices to carry a magnetic flux as well as the proton vortices.
The largest contribution to the pinning force is then due to the magnetic interaction
of the vortices, giving a pinning energy per site of
(e.g., see Mendell 1991a; Jones 1991; Chau, Cheng \& Ding 1992)
$$
\epsilon_{\rm pin} \approx
{1 \over 8 \pi}
\left ( {\rhonp \over \rhopp} \right )
\left ( {2 \Phi_o^2 {\rm cos}\theta_{np}
\over \pi \Lambda {\rm sin}\theta_{np}} \right )
\approx 10^{-4} {\rm ergs},  \eqno\stepeq
$$
where $\theta_{np}$ is the angle between the vortices.
For $d_n > d_p$ the number of pinning sites per unit volume is roughly
$$
N_{\rm pin} \approx {1 \over d_n^2 d_p} \approx {n_{{\rm v}n} \over d_p}. \eqno\stepeq
$$
The magnitude of the pinning force can be compared with the smaller of the vortex forces
(that of the neutron vortices) by considering the ratio of the energy density associated
with each force, i.e.,
$$
{f_{\rm pin} \over f_{{\rm {\small V}}n}}
\approx {\epsilon_{\rm pin} N_{\rm pin} \over \varepsilon_n n_{{\rm v}n}}
\approx 10^{-4} B_{o\stwelve}{}^{1/2}, \eqno\stepeq
$$
where $\varepsilon_n \approx 10^9 {\rm erg} \cdot {\rm cm}^{-1}$ has been
used (see Mendell 1991a). Thus pinning forces between the vortices will be
ignored in this paper.

Given the caveats of the last paragraph, expressions for
${\blambda}_n$ and ${\blambda}_p$ are now found by
constructing a model of the energy density.  The magnetic energy density is
complicated by the fact that part of the magnetic induction is confined
to the vortices, and also by the `drag' effect which causes the neutron
vortices to carry a magnetic flux, as noted above.  Following Alpar et al. (1984)
the magnetic induction due to the vortices is
$$
\bmath{B}_{\rm vortex} 
= \Phi_o n_{{\rm v}p}{\bnu}_p 
+ \left ( {m_p \over m_n} \right )
\left ( {\rhonp \over \rhopp} \right ) \Phi_o n_{{\rm v}n}{\bnu}_n . \eqno\stepeq
$$
Note that the first and second terms on the right side of this equation
correspond to the magnetic induction due to the proton and neutron vortices,
$\bmath{B}_{{\rm v}p}$ and $\bmath{B}_{{\rm v}n}$, respectively.

In addition to the magnetic induction of the vortices a non-quantized magnetic
field known as the London field, $\bmath{B}_{\rm {\small L}}$, also exists whenever the
vorticities of the fluids are nonzero (London 1960). Using equation (18),
the London equations, (9) and (10), and  $\bmath{B} = \bmath{B}_{\rm vortex}
+ \bmath{B}_{\rm {\small L}}$, gives
$$
\bmath{B}_{\rm {\small L}} = - {m_p c \over e} {\bomega}_p 
- {m_p c \over e}{\rhonp \over \rhopp}{\bomega}_n . \eqno\stepeq
$$
Ignoring interactions between the vortices, the vortex
energy density is just the vortex energy per unit length times the vortex density.
The energy density of the vortices and electro-magnetic field then becomes
$$
\eqalign{
& U_{{\rm {\small V}}} +  U_{{\rm {\small EM}}} = 
\varepsilon_p n_{{\rm v}p}
+ \varepsilon_n n_{{\rm v}n} \cr
& \qquad + {1 \over 8\pi} [ \vert \bmath{B}_{{\rm vortex}} + \bmath{B}_{\rm {\small L}} \vert^2 
-  B_{{\rm v}p}^2 - B_{{\rm v}n}^2 +  D^2] . \cr } \eqno\stepeq
$$
Note that the magnetic energy density of each vortex species is subtracted off since
it has already been accounted for in the energy per unit length of the vortices
[see eq. (4)].   It has also been assumed that the displacement
field $\bmath{D}$ equals the electric field $\bmath{E}$.  Noting that
electromagnetic energy density has the standard differential (e.g., see
Jackson 1975)
$$
{\bf d} U_{{\rm {\small EM}}}  = {1 \over 4 \pi} \bmath{E} \cdot {\bf d} \bmath{D}
+ {1 \over 4 \pi} \bmath{H} \cdot {\bf d} \bmath{B}  \eqno\stepeq
$$  
and using equations (13), (18)--(21), and the London equations,
it can be shown that $\bmath{H} = \bmath{B}$, $\bmath{E} = \bmath{D}$, and
$$
{\blambda}_n = {2m_n \over h}\varepsilon_n 
{{\bomega}_n \over \vert {\bomega}_n \vert}
- {1 \over 4\pi}
\Bigl ( {m_p c \over e} \Bigr )^2 \Bigl ( {\rhonp \over \rhopp} 
\Bigr)^2 {\bomega}_n ,  \eqno\stepeq
$$
$$
{\blambda}_p = {2m_p \over h}\varepsilon_p {{\bomega}_p +
a_p\bmath{B} \over \vert {\bomega}_p + a_p\bmath{B} \vert }
- {1 \over 4\pi} \Bigl ( {m_p c \over e} \Bigr )^2 
({\bomega}_p + a_p\bmath{B}) . \eqno\stepeq
$$
Similar results for ${\blambda}_n$ and ${\blambda}_p$ are given
in Mendell (1991a).\note{${}^\dagger$}{That paper includes the weak dependance
of $\varepsilon_n$ on $d_n$ (which partly measures the interaction between these
vortices) and subtracts $B_{\rm vortex}^2$ (which includes an interaction energy
between the vortices) rather than just $B_{{\rm v}p}^2 + B_{{\rm v}n}^2$ as in
equation (20).  Since interactions between vortices are ignored in this paper, these
differences are insignificant.}

The main purpose of this section is to present the equations for the neutron-star
outer core given in Mendell (1991a) in the MHD limit, including the effects of
the magnetic field and the vortices (which are ignored in the final set of
equations in that paper).  Here the very small fraction of muons are ignored.
Since the system is very degenerate the entropy density of the electrons
and any excitations of the neutrons and protons are also ignored, while
the largest of the dissipative terms found in Mendell (1991b) are included.
The equations are for small perturbations away from a stationary and
axisymmetric equilibrium. (This allows weak interactions to be ignored.)
The perturbed quantities are
prefixed with a $\delta$, and equilibrium quantities are given a subscript $o$.
The equilibrium velocity $\bmath{v}_o$ (vorticity ${\bomega_o}$) is assumed
to be that of uniform rotation with constant angular velocity while the
equilibrium magnetic induction is taken to be an arbitrary stationary
vector field $\bmath{B}_o$.

The MHD equations then are as follows:  the conservation laws for the 
neutron and proton mass densities $\rhon$ and $\rhop$ [note the appearance of
the mass density matrix, given in eqs. (11) and (12)], given by
$$
\partial_t \delta \rhon + \bmath{v}_o \cdot {\bnabla} \delta \rhon
+ {\bnabla}\cdot ( \rhonn \delta \bmath{v}_n 
+ \rhonp \delta \bmath{v}_p) = 0 , \eqno\stepeq
$$
$$
\partial_{t} \delta \rhop + \bmath{v}_o \cdot {\bnabla} \delta \rhop
+ {\bnabla}\cdot (\rhonp \delta \bmath{v}_n 
+ \rhopp \delta \bmath{v}_p) = 0,\eqno\stepeq
$$
Maxwell's equations for the
electric field $\bmath{E}$, magnetic induction $\bmath{B}$, charge density
$\sigma$, and current density $\bmath{J}$, given by
$$
\delta \sigma = {e \rhop \over  m_p} - {e \rhoe \over  m_e} = 0,  \eqno\stepeq
$$
$$
{\bnabla} \cdot \delta \bmath{E} = 0, \eqno\stepeq
$$
$$
{\bnabla} \cdot \delta \bmath{B} = 0, \eqno\stepeq
$$
$$
{\bnabla} \times \delta \bmath{E} = -{1\over c}\partial_t \delta \bmath{B}, \eqno\stepeq
$$
$$
{\bnabla} \times \delta \bmath{B} = {4\pi \over c} \delta \bmath{J}, \eqno\stepeq
$$
$$
\eqalign{
\delta \bmath{E} = & - {\delta \bmath{v}_e \over c} \times \bmath{B}_o
- {\bmath{v}_o \over c} \times \delta \bmath{B} \cr
& - {m_p \over e} \left ( {\rhon \over \rhop} \delta \bmath{F}_{n({\rm mf})}
+ \delta \bmath{F}_{p({\rm mf})} \right ) , \cr } \eqno\stepeq
$$
Newton's 2nd law for the neutron superfluid velocity, $\bmath{v}_n$, and the 
average velocity of the charged fluids, $\bmath{u}$, given by 
$$
\eqalign{
\partial_t \delta \bmath{v}_n & = - \bmath{v}_o \cdot {\bnabla} \delta
\bmath{v}_n - \delta \bmath{v}_n \cdot {\bnabla} \bmath{v}_o   
- {\bnabla} \delta \mu_n  \cr
& + \rpn (\delta \bmath{v}_p - \delta \bmath{v}_n) \times {\bomega}_o
+ \delta \left [ {1 \over \rhon} ({\bnabla} \times {\blambda}_n)
\times {\bomega}_n  \right ] \cr
& - {\bnabla} \delta \varphi + \delta \bmath{F}_{n({\rm mf})} , \cr} \eqno\stepeq
$$
$$
\eqalignno {
\partial_t \delta \bmath{u} & = - \bmath{v}_o \cdot {\bnabla} \delta \bmath{u}
- \delta \bmath{u} \cdot {\bnabla} \bmath{v}_o
- {\rhop \over \tilde{\rhop}} {\bnabla} \delta \mu_p
- {\rhoe \over \tilde{\rhop}} {\bnabla} \delta \mu_e & \startsubeq \cr
& +  {1 \over \tilde{\rhop}}
\left ( {\delta \bmath{J} \over c} \times \bmath{B}_o \right )
+ {\rhonp \over \tilde{\rhop}}(\delta \bmath{v}_n - \delta \bmath{v}_p)
\times {\bomega}_o & \stepsubeq \cr
& + {\rhop \over \tilde{\rhop}}
\delta \left [ {1 \over \rhop}( {\bnabla} \times {\blambda}_p)
\times ({\bomega}_p + a_p \bmath{B}) \right ] & \stepsubeq \cr
& - {\bnabla} \delta \varphi
- {\rhon \over \tilde{\rhop}} \delta \bmath{F}_{n({\rm mf})}
- {1 \over \tilde{\rhop}} {\bnabla} \cdot {\btau} , & \stepsubeq \cr
} 
$$
and Newton's law of gravitation for the gravitational potential $\varphi$, given by
$$
\nabla^2 \delta \varphi = 4 \pi G \delta \rho . \eqno\stepeq
$$

Consider equations (26) and (27).  These equations follow
from the MHD assumption that charge neutrality is preserved, and
equation (26) also
relates $\rhoe$ to $\rhop$.  Equations (28) and (29) 
are just the usual Maxwell's equations, followed by 
equation (30) which is Ampere's law ignoring the displacement current term.
This is justified for waves with frequency $\omega$ and wave vector $k$
when $(\omega / ck)^2 \ll 1$, and this corresponds to the
second part of the MHD limit.  Next, equation (31) follows from considering
the dynamics of the current density and is valid for frequencies small compared
to the plasma frequency and in the limit of large conductivity.  Note that this
equation implies that the Lorentz force on the electrons is
vanishingly small in the MHD limit.  Also in this equation, 
$\bmath{F}_{n({\rm mf})}$ and $\bmath{F}_{p({\rm mf})}$ are the
mutual friction forces due to electron scattering off the neutron
and proton vortice respectively.  These terms are
small, though they are the largest of the dissipative corrections
to this equation.  Dissipative effects are discussed in Section 5.

Now consider equations (32) and (33). 
In these equations the $\mu$'s are the chemical potentials
defined in equation (33) in Mendell (1991a).  Equation (32) is just the usual
Landau equation for a superfluid modified by the `drag' effect and the 
vortex force of the neutron vortices [see eqs. (9) and (14)]. Equation (33)
is for the average velocity of the charged fluid defined by
$$
\bmath{u} \equiv {\rhop \over \tilde{\rhop}}\bmath{v}_p + {\rhoe
\over \tilde{\rhop}} \bmath{v}_e , \eqno\stepeq
$$
where $\tilde{\rhop} = \rhop + \rhoe$,  and it is the standard
MHD equation for this quantity modified by the `drag effect'
and the vortex force of the proton vortices [see eqs. (10) and (14)].
Note that in equation (33) terms proportional to the equilibrium
charge density, which is exceedingly small and in fact zero in
non-rotating neutron stars, have been ignored [see eqs. (39) and (40)
in Mendell 1991a].  The viscous stress tensor, ${\btau}$, and
the mutual friction forces that appear in these equations are
considered further in Section 5.

One great simplification of the MHD limit is that the vector fields
$\bmath{J}$, $\bmath{E}$, $\bmath{v}_p$, and $\bmath{v}_e$, are easily
expressed in terms of $\bmath{B}$, $\bmath{u}$, and $\bmath{v}_n$.  Equations
(30) and (31) already specify $\bmath{J}$ and $\bmath{E}$ and it can be shown
that the velocities of the electrons and protons are given by  
$$
\delta \bmath{v}_e = \rpp \delta \bmath{u}
+ {\rhonp \over \rhop} \delta \bmath{v}_n - {m_p \over e \rhop} \delta \bmath{J}
, \eqno\stepeq
$$
$$
\delta \bmath{v}_p = \delta \bmath{u} - {\rhoe \over \rhop}
{\rhonp \over \rhop} \delta \bmath{v}_n + {m_p \rhoe \over e \rhop^2} \delta \bmath{J}
, \eqno\stepeq
$$
ignoring factors of $\rhoe / \rhop$ in these equations only for the moment.

To finish this section it will be useful to review
several of the characteristic frequencies associated
with charged particles and electromagnetic fields in terms
of the angular frequency $\omega$, the wavevector
$k = (2\pi /\lambda)$, and the wavelength $\lambda$ associated
with plane wave solutions.  

Two frequencies well known in plasma physics are the plasma
frequency, $\omega_{{\rm {\small P}}}$, and
the cyclotron frequency, $\Omega_{{\rm {\small C}}}$.
For motion of the protons these frequencies are given by
$$
\omega_{{\rm {\small P}}}
= \Bigl ( {4 \pi e^2 \rhop \over m_p^2} \Bigr )^{1/2}
\approx 3.2 \times 10^{21}
\rho_{p\sthirteen}{}^{1/2} \;
{\rm s}^{-1},\eqno\stepeq
$$
$$
\Omega_{{\rm {\small C}}} = {e B_o \over m_p c} \approx 9.6 \times 10^{15}
B_{o\stwelve}  \; {\rm s}^{-1}. \eqno\stepeq
$$
A characteristic frequency
associated with charge currents in metals and
superconductors is the helicon frequency,
$\omega_{\rm h}$, given by (see de Gennes \& Matricon 1964; 
Abrikosov, Kemoklidze \& Khalatnikov 1965)
$$
\omega_{\rm h} = \Omega_{{\rm {\small C}}}
{c^2 k^2 \over \omega_{{\rm {\small P}}}^2}
\approx 3.3 \times 10^{-17}
\rho_{p\sthirteen}{}^{-1}
\lambda_{\ssix}{}^{-2}
B_{o\stwelve} 
 \; {\rm s}^{-1} , \eqno\stepeq
$$
where $\lambda_{\ssix}$ is $\lambda$ divided by $10^6 \; {\rm cm}$.

Of main interest in this paper are the Alfv\'{e}n waves found in the
MHD limit of ordinary plasma physics. For an ordinary proton-electron
plasma these waves travel along magnetic field lines with a frequency given by
$$
\omega_{{\rm {\small A}}p} =  \sqrt{ \Omega_{\rm {\small C}} \omega_{\rm h}}
= \sqrt{{B_o^2 / 4 \pi \over \rhop}} \; k
= {B_o \over \sqrt{4 \pi \rhop}} k . \eqno\stepeq
$$
These waves can be compared
with those of a vibrating string with the tension of the field lines provided
by magnetic energy density ($\sim B_o^2/4\pi$) and the inertia provided by
the mass that moves with the lines ($\rhop$).
In the case of an ordinary-fluid neutron star the neutrons and protons would
act as a single fluid and the density $\rhop$ would be replaced with the 
total mass density $\rho$, giving 
$$
\omega_{\rm {\small A}} = {B_o \over \sqrt{4 \pi \rho}} k 
\approx 0.056
\rho_{\sfifteen}{}^{-1/2}
\lambda_{\ssix}{}^{-1}
B_{o\stwelve}
\; {\rm s}^{-1} , \eqno\stepeq
$$
where $\rho_{\sfifteen}$ is the density divided by
$10^{15} \; {\rm g} \cdot {\rm cm}^{-3}$.
It will shown in the next section that Alfv\'{e}n waves 
do not occur in the exotic superfluid plasma considered here.  

There are also several characteristic frequencies associated with
oscillations of the underlying vortices. (See footnote ${}^\star$;
Krusius, et al. 1993; and Sonin 1987 for studies and reviews
of vortex oscillations in superfluids; see Friedel, de Gennes
\& Matricon 1963; Tsui, et al. 1994;
Kopnin, et al. 1995; Blatter \& Ivlev 1995;
and Sonin 1996a,b for studies of vortex oscillations in superconductors.)
The frequency of circularly polarized vortex oscillations which
travel parallel to the vortices, known as Kelvin waves, are given by
$$
\omega_{{\rm {\small V}}n} = {2 m_n \over h} {\varepsilon_n k^2 \over \rhon}
\approx 2.0 \times 10^{-14}
\varepsilon_{n\snine}
\rho_{n\sfifteen}{}^{-1}
\lambda_{\ssix}{}^{-2}
\; {\rm s}^{-1} , \eqno\stepeq
$$
$$
\omega_{{\rm {\small V}}p} = {2 m_p \over h}{\varepsilon_p k^2 \over \rhop}
\approx 2.0 \times 10^{-15}
\varepsilon_{p\ssix} 
\rho_{p\sthirteen}{}^{-1}
\lambda_{\ssix}{}^{-2}
\; {\rm s}^{-1} , \eqno\stepeq
$$
for the neutron and proton vortices, respectively.
In these equations $\rho_{n\sfifteen}$ is $\rhon$ divided by
$10^{15} \; {\rm g} \cdot {\rm cm}^{-3}$, and
$\varepsilon_{n\snine}$ and $\varepsilon_{p\ssix}$ are
$\varepsilon_n$ and $\varepsilon_p$ divided by
$10^9 \; {\rm erg} \cdot {\rm cm}^{-1}$ and
$10^6 \; {\rm erg} \cdot {\rm cm}^{-1}$ respectively.
These vortex frequencies can be shown to arise from the first
terms in equations (22) and (23) for ${\blambda}_n$ and 
${\blambda}_p$. 

\section{Alfv\'{e}n and Cyclotron-vortex waves}

In this section is it shown that Alfv\'{e}n waves, which
occur in ordinary plasmas, do not occur in the exotic plasma
considered in this paper.  Instead a new frequency is shown to 
exist that depends on the cyclotron and vortex frequencies
of the fluid. Thus it is referred to as the cyclotron-vortex frequency. 
However these waves are analogous to Alfv\'{e}n waves
with the tension due to the magnetic energy density replaced
by the vortex energy density. A straightforward derivation of
this frequency is given below.

To begin with, consider uniform neutron-star matter. 
For simplicity assume that the matter is not rotating and that there
is a uniform equilibrium magnetic induction which defines the
$z$-axis, i.e., take $\Omega = 0$ and $\bmath{B} = B_o \hat{\bmath{z}}$.
Next, look for plane wave solutions with the wave vector parallel
to the $z$-axis, i.e., let $\bmath{k} = k\hat{\bmath{z}}$. The
space-time dependence of all the perturbed quantities can then be taken
to be completely specified by ${\rm exp}(ikz -i\omega t)$.
Making these substitutions the equations in Section 2 become purely algebraic. 
The equations also decouple into components that are parallel and
transverse to the $z$-axis.  Ignoring dissipative effects, which will
be considered in Section 5, the resulting equations for the transverse
components are
$$
ik(\hat{\bmath{z}} \times \delta \bmath{E}) = {i \omega \over c} \delta \bmath{B} ,
\eqno\stepeq
$$
$$
ik(\hat{\bmath{z}} \times \delta \bmath{B}) = {4 \pi \over c} \delta \bmath{J} ,
\eqno\stepeq
$$
$$
\delta \bmath{E} = -({\delta \bmath{v}_{e} \over c} \times B_o \hat{\bmath{z}}) ,
\eqno\stepeq
$$
$$
\delta \bmath{v}_n = 0 , \eqno\stepeq
$$
$$
-i \omega \delta \bmath{u} = {i k \over \tilde{\rhop}}[(\hat{\bmath{z}} \times
\delta {\blambda}_p) \times a_p B_o \hat{\bmath{z}} ] + {1 \over \tilde{\rhop}}
({\delta \bmath{J} \over c} \times B_o \hat{\bmath{z}}) . \eqno\stepeq
$$

A few steps will now show that Alfv\'{e}n waves do not exist
in the solution to these equations.  Using equation (46) it is easy
to show that
$$
\delta \bmath{B} = {4 \pi \over ikc}(\delta \bmath{J} \times \hat{\bmath{z}}). \eqno\stepeq
$$
Then consider the vortex force acting on $\delta \bmath{u}$.  Equation (23)
implies that $\delta {\blambda}_p$ is given by
$$
\delta {\blambda}_p= \left [ {2 m_p \over h}
{\varepsilon_p \over \Omega_{{\rm {\small C}}}} -
{m_p^2 c^2 \over 4 \pi e^2} \right ]
[ik(\hat{\bmath{z}} \times \delta \bmath{v}_p) + a_p \delta \bmath{B}]. \eqno\stepeq
$$
Using equation (50) the vortex force in equation (49) can be simplified to become
$$
\eqalign {
& {i k \over \tilde{\rhop}}[(\hat{\bmath{z}} \times
\delta {\blambda}_p) \times a_p B_o \hat{\bmath{z}}] \, =  \, {\rhop \over
\tilde{\rhop}} (\omega_{{\rm {\small V}}p}
- \omega_{{\rm h}})(\delta{v}_p \times \hat{\bmath{z}}) \cr
& \qquad + {1 \over \tilde{\rhop}}{\omega_{{\rm {\small V}}p}
\over \Omega_{{\rm {\small C}}}} {\omega_{{\rm {\small P}}}^2 \over
c^2 k^2} ( {\delta \bmath{J} \over c} \times B_o \hat{\bmath{z}})
- {1 \over \tilde{\rhop}}( {\delta \bmath{J} \over c} \times B_o
\hat{\bmath{z}}) . \cr } \eqno\stepeq
$$
Note that the last term on the right side of equation (52), which arises from
the interaction between the vortex and London fields, will cancel the
last term in equation (49) [same as the first term in eq. (33b)].
Since this is the term that usually gives rise to Alfv\'{e}n waves
these waves do not occur in the superconducting plasma considered here.
It also should be clear that this result does not depend on the specifics
of this plasma but should be true for any plasma with one component
superconducting and in the vortex state. Furthermore this result applies
not just to axial modes.  In general the last term in equation (51) will
give a force equal to
$(-1/4\pi\tilde{\rhop})({\bnabla} \times \delta \bmath{B}) \times \bmath{B}_o$ which
will give the same cancellation.   

From here on terms smaller by a factor $\rhoe / \rhop$ will be dropped in all equations.
Equation (36) already uses this approximation, and substituting this into
equation (47) gives
$$
\delta \bmath{E} = -{\rhopp \over \rhop}
\left ( {\delta \bmath{u} \over c} \times B_o \hat{\bmath{z}} \right )
+ {m_p \over e \rhop}
({\delta \bmath{J} \over c} \times B_o \hat{\bmath{z}}) , \eqno\stepeq
$$
while equations (45) and (46) give
$$
\delta \bmath{E} = {4 \pi i \omega \over c^2 k^2} \delta \bmath{J} . \eqno\stepeq
$$
From these one can show
$$
\rhopp(\delta \bmath{u} \times \hat{\bmath{z}}) = 
- {\omega_{{\rm {\small P}}}^2
\over c^2 k^2}{i \omega \over
\Omega_{{\rm {\small C}}}^2} {\delta \bmath{J} \over c} B_o
+ {1 \over \Omega_{{\rm {\small C}}}}
({\delta \bmath{J} \over c} \times B_o \hat{\bmath{z}}) . \eqno\stepeq
$$

Substituting equations (37), (52), and (55) into equation (49) and
simplifying results in the following eigenvalue equation:
$$
\eqalign {
& \left [\omega^2 - \omega_{{\rm h}}(\omega_{{\rm {\small V}}p}
- \omega_{{\rm h}}) - {\rhopp \over \rhop} \Omega_{{\rm {\small C}}}
\omega_{{\rm {\small V}}p} \right ](\delta \bmath{J} \times \hat{\bmath{z}}) \cr
& \qquad + i \omega (\omega_{{\rm {\small V}}p} 
- 2 \omega_{{\rm h}}) \delta \bmath{J} = 0 . \cr } \eqno\stepeq
$$
Solving equation (56), the eigenvalues are given by 
$$
\omega^2 - \omega_{{\rm h}}(\omega_{{\rm {\small V}}p} - \omega_{{\rm h}})
- {\rhopp \over \rhop} \Omega_{{\rm {\small C}}} \omega_{{\rm {\small V}}p}
= \pm \omega (\omega_{{\rm {\small V}}p} - 2 \omega_{{\rm h}})
, \eqno\stepeq
$$
and the components of the eigenvectors are related by
$\delta J_x = \mp i \delta J_y$.  The two sign choices in equation (57)
give eigenfrequencies
$$
\omega = \left \{ \eqalign { & \pm \sqrt{{\rhopp \over \rhop}
\Omega_{{\rm {\small C}}} \omega_{{\rm {\small V}}p}} 
- {\omega_{{\rm {\small V}}p} \over 2} + \omega_{{\rm h}}  \cr
& \pm \sqrt{{\rhopp \over \rhop}
\Omega_{{\rm {\small C}}} \omega_{{\rm {\small V}}p}} 
+ {\omega_{{\rm {\small V}}p} \over 2} - \omega_{{\rm h}}  \cr } \right. 
\eqno\stepeq
$$
where terms smaller than those kept by at least a factor of $\omega_{{\rm
{\small V}}p}/\Omega_{{\rm {\small C}}}$ have been dropped.
As far as this author knows the first term in equation (58) has not been
derived previously from MHD theory.  Since it
depends on the cyclotron frequency $\Omega_{{\rm {\small C}}}$ and the
Kelvin vortex oscillation frequency $\omega_{{\rm {\small V}}p}$ it is natural
to call it the cyclotron-vortex frequency, $\omega_{{\rm {\small CV}}}$.
In a neutron star it will have a magnitude given by
$$
\eqalign {
\omega_{{\rm {\small CV}}} & = 
\sqrt { {\rhopp \over \rhop} \Omega_{{\rm {\small C}}} \omega_{{\rm {\small V}}p} }
 = \sqrt{ \left ( {\rhopp \over \rhop} \right )
\left ( {\varepsilon_p \over \rhop} \right )
\left ( {B_o \over \Phi_o} \right ) } \; k \cr
& \approx 6.4
\lambda_{\ssix}{}^{-1}
\left ( {\rhopp \over \rhop} \right )
\sqrt {(1 + \Delta) B_{o\stwelve}}
\; {\rm s}^{-1} . \cr }  \eqno\stepeq
$$
Note that $\omega_{{\rm {\small CV}}}$ is quite different from the
Alfv\'{e}n frequency since it is nearly independent of the mass density
(except through $\Delta$) and depends on $\sqrt{B_o}$.  However, consider
the definition of the critical field
$H_{c1} = (\Phi_o /4 \pi \Lambda^2) \ln (\Lambda / \xi)$,
above which the vortex state is favored in a type II superconductor
(and approximately the field within the London depth of one proton vortex).
Noting that $\varepsilon_p = \Phi_o H_{c1} / 4 \pi$ and
$n_{{\rm {\small V}}p} = B_o/\Phi_o$ then
$$
\omega_{{\rm {\small CV}}}
\approx
\sqrt{{\varepsilon_p n_{{\rm {\small V}}p} \over \rhop}} \; k 
= \sqrt{{H_{c1}B_o / 4 \pi \over \rhop}} \; k . \eqno\stepeq
$$
It can seen then that $\omega_{{\rm {\small CV}}}$
is analogous to the Alfv\'{e}n frequency with the 
tension due to the magnetic energy density replaced by
the vortex energy density $\varepsilon_p n_{{\rm {\small V}}p}$.
This equals $H_{c1}B_o / 4 \pi$ (roughly the magnetic energy density
of the vortex array).  Easson \& Pethick (1977) found this same
factor in their study of the stress tensor for a type II superconductor,
though they did not consider its implication on the dynamics.

\section{Including Rotation}

To see how the modes in the rotating case are affected by the cyclotron-vortex frequency,
the solutions of the last section are extended here to the case of uniform rotation.
Let the rotation axis define the $z$-axis of the system. 
To find plane wave solutions first transform to
coordinates which corotate with the fluid, given by
$$
x' = x \cos (\Omega t) + y \sin (\Omega t) ,\eqno\stepeq
$$
$$
y' = - x \sin (\Omega t) + y \cos (\Omega t) , \eqno\stepeq
$$
$$
z' = z , \eqno\stepeq
$$
$$
t' = t . \eqno\stepeq
$$
The spatial derivative operator ${\bnabla}$ is covariant under this transformation, 
whereas the convective derivative of any fluid velocity $\bmath{V}$ transforms as
$$
\partial_t \bmath{V} + \bmath{v}_o \cdot {\bnabla} \bmath{V}
+ \bmath{V} \cdot {\bnabla} \bmath{v}_o
= \partial_{t'} \bmath{V}' - 2 \Omega (\bmath{V}' \times \hat{\bmath{z}}) . \eqno\stepeq
$$
If we also define new electric and magnetic fields
$$
\bmath{E}' =  \bmath{E} + {\bmath{v}_o \over c} \times \bmath{B}, \eqno\stepeq
$$
$$
\bmath{B}' = \bmath{B} , \eqno\stepeq
$$
and ignore corrections of order $(v_o/c)(\omega/ck) \ll 1$ then Maxwell's equations
remain covariant under these transformations.

In accordance with the last section we consider axial modes
with spacetime dependence ${\rm exp} (i kz' - i \omega t')$
and for simplicity assume the equilibrium magnetic field is aligned with the rotation axis.
After making the above transformations and dropping all primes one finds that Maxwell's
equations remain identical to those in the last section, and that the equations
for the fluid velocities become
$$
- {i \omega \over 2 \Omega} \delta \bmath{v}_n  - (\delta \bmath{v}_n \times \hat{\bmath{z}}) =
{\rhonp \over \rhon}( \delta \bmath{v}_p - \delta \bmath{n}_n) \times \hat{\bmath{z}}
+ {i k \over \rhon } \delta {\blambda}_n
, \eqno\stepeq
$$
$$
\eqalign {
& - {i \omega \over 2 \Omega} \delta \bmath{u}
- (\delta \bmath{u} \times \hat{\bmath{z}}) =
{\rhonp \over \rhop}(\delta \bmath{v}_n - 
\delta \bmath{v}_p) \times \hat{\bmath{z}} \cr
& \qquad + {(2 \Omega + \Omega_{\rm {\small C}})i k \over 2\Omega \rhop }
\delta {\blambda}_p + {1 \over 2\Omega \rhop}
({\delta \bmath{J} \over c} \times B_o \hat{\bmath{z}}) , \cr }
\eqno\stepeq
$$
where
$$
\delta {\blambda}_n = \left [ {2 m_n \varepsilon_n \over 2 \Omega h } -
\left ( {\rhonp \over \rhopp} \right )^2 {m_p^2 c^2 \over 4 \pi e^2} \right ]
(ik\hat{\bmath{z}} \times \delta \bmath{v}_n)
, \eqno\stepeq
$$
$$
\delta {\blambda}_p = 
\left [ {2 m_p \varepsilon_p \over h (2 \Omega + \Omega_{{\rm {\small C}}}) } -
{m_p^2 c^2 \over 4 \pi e^2} \right ]
[ik(\hat{\bmath{z}} \times \delta \bmath{v}_p) + a_p \delta \bmath{B}]
. \eqno\stepeq
$$

The goal will be to express everything in terms of $\delta \bmath{v}_n$ and $\delta \bmath{u}$.
Using equations (31) and (54) one can show
$$
\delta \bmath{v}_e = \left ( {m_p \over e \rhop} \right )
\left ( { i \omega \over \omega_{\rm h}} \right )
(\delta \bmath{J} \times \hat{\bmath{z}}) . \eqno\stepeq
$$
Substituting into equation (36) gives
$$
\delta \bmath{J} + {i \omega \over \omega_{\rm h}}(\delta \bmath{J} \times \hat{\bmath{z}})
= {e \rhop \over m_p} \left ( \rpp \delta \bmath{u}
+ {\rhonp \over \rhop} \delta \bmath{v}_n \right) .\eqno\stepeq
$$
For $\omega \gg \omega_{\rm h}$ the following approximations can be made:
$$
\delta \bmath{J} \times \hat{\bmath{z}}
\approx {\omega_{\rm h} \over i \omega}
{e \rhop \over m_p} \left ( \rpp \delta \bmath{u}
+ {\rhonp \over \rhop} \delta \bmath{v}_n \right) , \eqno\stepeq
$$
and from equations (37) and (50)
$$
\delta \bmath{v}_p \approx \delta \bmath{u}
- {\rhoe \rhonp \over \rhop^2} \delta \bmath{v}_n, \eqno\stepeq
$$
$$
a_p \delta \bmath{B} \approx - \left ( {\Omega_{\rm {\small C}} k \over \omega} \right )
\left ( \rpp \delta \bmath{u} + {\rhonp \over \rhop} \delta \bmath{v}_n \right) .\eqno\stepeq
$$
Substituting these into equations (68) - (71) and simplifying gives
$$
\eqalign {
& i \omega \delta \bmath{v}_n
+ 2 \Omega {\rhonp \over \rhon}(\delta \bmath{u} \times \hat{\bmath{z}}) \cr
& + 2 \Omega \left ( {\rhonn \over \rhon}
+ {\omega_{{\rm {\small V}} n} \over 2 \Omega} 
- {c^2 k^2 \over \omega_{\rm {\small P}}^2}{\rhop \rhonp^2 \over \rhon \rhopp^2}
\right ) (\delta \bmath{v}_n \times \hat{\bmath{z}}) = 0 , \cr }
\eqno\stepeq
$$
$$
\eqalign{
& 2 \Omega {\rhonp \over \rhop} (\delta \bmath{v}_n \times \hat{\bmath{z}})
+ {\rhonp \over \rhop} \left ( {2 i \Omega \omega_{\rm h} \over \omega}
 - {i \Omega_{\rm {\small C}} \omega_{{\rm {\small V}}p} \over \omega}
\right ) \delta \bmath{v}_n \cr
& + 2 \Omega \left ( {\rhopp \over \rhop}
+ { \omega_{{\rm {\small V}}p} \over 2 \Omega}
- {c^2 k^2 \over \omega_{\rm {\small P}}^2} - {\omega_{\rm h} \over 2 \Omega}
\right ) (\delta \bmath{u} \times \hat{\bmath{z}}) \cr
& + \left ( i \omega
- {i \Omega_{\rm {\small C}} \omega_{{\rm {\small V}}p} \over \omega}
{\rhopp \over \rhop}
+ {2 i \Omega \omega_{\rm h} \over \omega}{\rhopp \over \rhop}
\right ) \delta \bmath{u} = 0 .\cr
}  \eqno\stepeq
$$

To solve these equations one can invert equation (77) to solve for
$\delta \bmath{u}$ and then substitute into equation (78) to get a simple
eigenvalue equation for $\delta \bmath{v}_n$.  
For $\omega$ , $\Omega \gg \omega_{\rm h}$, $\omega_{{\rm {\small V}}p}$,
$\omega_{{\rm {\small V}}n}$, and $c^2k^2 \ll \omega_{\rm {\small P}}^2$ the
resulting dispersion relationship reduces down to
$$
\eqalign{
& \omega^3
\pm 2 \Omega \left ( {\rhonn \over \rhon} + {\rhopp \over \rhop} \right) \omega^2 \cr
& + \left ( 4 \Omega^2 {\varrho^2 \over \rhon \rhop} 
- \Omega_{\rm {\small C}}
\omega_{{\rm {\small V}}p}{\rhopp \over \rhop} \right ) \omega
\mp 2 \Omega \Omega_{\rm {\small C}} \omega_{{\rm {\small V}}p}
{\varrho^2 \over \rhon \rhop} = 0 . \cr }
\eqno\stepeq
$$
where $\varrho^2 = \rhonn\rhopp - \rhonp^2$.
The eigenvectors are related by
$\delta \bmath{u}_x = \pm i \delta \bmath{u}_y$ and
$$
\delta \bmath{u} = 
{\rhon \over \rhonp}\left ( \pm {\omega \over 2 \Omega}
 - {\rhonn \over \rhon} \right ) \delta \bmath{v}_n .  \eqno\stepeq
$$
In the limit $B_o = 0$ equation (79) can be solved exactly to give    
$$
\omega = \pm \left \{ \eqalign {
& 2\Omega \cr
& 2\Omega {\varrho^2 \over \rhon \rhop} \cr
 } \right. \eqno\stepeq
$$
where the upper branch is usually called the inertial mode
and the lower branch is a new mode that appears due to the `drag' effect.
Keeping $B_o \ne 0$ an approximate solution to equation
(79) can be found by noting that $\rhop \ll \rhon$, $\rhonn \approx \rhon$,
and $\varrho^2 \approx \rhon \rhopp$. The solution then is
$$
\omega = \pm \left \{ \eqalign {
& 2\Omega \cr
& \Omega {\rhopp \over \rhop}
 \pm \sqrt{ {\rhopp \over \rhop} \Omega_{\rm {\small C}} \omega_{{\rm {\small V}}p}
 + \Omega^2 \left ( {\rhopp \over \rhop} \right )^2 } \cr 
 } \right. \eqno\stepeq
$$
Note that the cyclotron-vortex frequency will dominate the effects
of rotation only when
$$
B_o \ge \left ( {\rhopp \over \rhop} \right ) \left ( {m_p c \over e} \right )
\left ( {\Omega^2 \over \omega_{{\rm {\small V}}p}} \right )
\approx {2.4 \times 10^{14} \over 1 + \Delta}
\Omega_{\stwo}{}^2
\lambda_{\ssix}{}^2
\; {\rm G}. \eqno\stepeq
$$ 
Thus this frequency will dominate the effects of rotation only in 
slowly rotating neutron stars with strong interior magnetic fields.
In the more usual case we can expand the lower branch in equation (82)
in powers of $\omega_{{\rm {\small CV}}}/ 2\Omega$.  Keeping only the lowest order
terms, the result is
$$
\omega = \pm \left \{ \eqalign {
& 2\Omega \cr
& 2\Omega {\rhopp \over \rhop} \cr 
& {\omega_{{\rm {\small CV}}}^2 \over 2 \Omega} {\rhop \over \rhopp} \cr
 } \right. \eqno\stepeq
$$
The top and middle branch are as in equation (81), while the bottom branch is
a new low frequency mode that depends on $\omega_{{\rm {\small CV}}}$.
Finally, recall that the frequencies in these equations are for the rotating frame; one
must add $\mp \Omega$ to get the frequencies in the inertial frame.

\section{Simplified Equations and Dissipative Effects}

It has been shown in the last two sections that of the vortex and magnetic forces the
dominant term gives rise to cyclotron-vortex waves that replace Alfv\'{e}n waves.
The next largest terms (e.g., consider $\omega_{{\rm {\small V}}p}
/ \omega_{{\rm {\small CV}}}$) are roughly $15$ orders of magnitude smaller for
typical neutron star numbers. This suggests dropping the vortex force of the neutrons
and approximating $\delta {\blambda}_p$ as
$$
\delta {\blambda}_p \approx {2 m_p \over h}\varepsilon_p
 {\delta \bmath{B} \over \vert \bmath{B}_o \vert } . \eqno\stepeq
$$
It also will be convenient to assume that terms proportional to
${\bnabla}(\varepsilon_p / \vert \bmath{B}_o \vert)$ are small and can be dropped.  Terms
smaller by a factor of $\rhoe/\rhop$ than those kept will also be ignored.
Furthermore, equations (29) and (31) can
be combined to eliminate $\delta \bmath{E}$ from the equations, and equations (36) and
(37) can be used to eliminate $\delta \bmath{v}_e$ and $\delta \bmath{v}_p$.  Finally, the
small corrections to the velocity fields proportional to $\delta \bmath{J}$ will also
be dropped.  With these approximations, equations (29) -- (33)
can be replaced by the following greatly simplified set of equations:
$$
\eqalign{
\partial_t \delta \bmath{v}_n = & - \bmath{v}_o \cdot {\bnabla} \delta
\bmath{v}_n - \delta \bmath{v}_n \cdot {\bnabla} \bmath{v}_o   
- {\bnabla}( \delta \mu_n + \delta \varphi) \cr
& + \rpn (\delta \bmath{u} - \delta \bmath{v}_n) \times {\bomega}_o
+ \delta \bmath{F}_{n({\rm mf})} , \cr} \eqno\stepeq
$$
$$
\eqalign {
\partial_t \delta \bmath{u} = & - \bmath{v}_o \cdot {\bnabla} \delta \bmath{u}
- \delta \bmath{u} \cdot {\bnabla} \bmath{v}_o
- {\bnabla}( \delta \mu_p + {m_e \over m_p} \delta \mu_e) \cr
& - {\bnabla} \delta \varphi + {\rhonp \over \rhop}
(\delta \bmath{v}_n - \delta \bmath{u}) \times {\bomega}_o \cr
& + {\varepsilon_p \over \rhop \Phi_o \vert \bmath{B}_o \vert}
({\bnabla} \times \delta \bmath{B}) \times \bmath{B}_o \cr
& - {\rhon \over \rhop} \delta \bmath{F}_{n({\rm mf})}
- {1 \over \rhop} {\bnabla} \cdot {\btau} , \cr} \eqno\stepeq
$$
$$
\eqalign {
\partial_t \delta \bmath{B} = {\bnabla} \times \Biggl [ &
({\rhopp \over \rhop} \delta \bmath{u}
+ {\rhonp \over \rhop} \delta \bmath{v}_n) \times \bmath{B}_o
+ \bmath{v}_o \times \delta \bmath{B} \cr
& \,\, + {1 \over a_p} \left ( {\rhon \over \rhop} \delta \bmath{F}_{n({\rm mf})}
+ \delta \bmath{F}_{p({\rm mf})} \right ) \Biggr ] , \cr } \eqno\stepeq
$$
These equations are in a useful form for studing the effects of the interior
magnetic field on the core spin-up during a pulsar glitch.  More will be said
about this in Section 6.

The remainder of this section will discuss the dissipative terms in these
simplified equations.  As discussed in Mendell (1991b) mutual friction due to 
electron-vortex scattering and the electron shear viscosity are the dominant
dissipative effects (see also Cutler \& Lindblom 1987; Cutler, Lindblom \&
Splinter 1990; Ipser \& Lindblom 1991; Lindblom \& Mendell 1995).
The viscous shear tensor is given by
$\delta {\btau} = - \eta_e \delta {\bf \Theta}$, where $\eta_e$ is the
electron shear viscosity approximately given by
$$
\eta_e = 6.0 \times 10^{22}
\rho_{\sfifteen}{}^2
T_{e\sseven}{}^{-2}
\; {\rm g}\cdot{\rm cm}^{-1}\cdot{\rm s}^{-1} 
, \eqno\stepeq
$$
where $T_{e\sseven}$ is the temperature of the electrons 
divided by $10^7 \; {\rm K}$ and
$\delta {\bf \Theta}$ is the shear tensor, given by
$$
\delta \Theta^a{}_b = {\scriptstyle {1 \over 2}} ( \nabla^a \delta v_{eb}
+ \nabla_b \delta v_e^a - {\scriptstyle {2 \over 3}} \delta^a_b
\nabla_c \delta v_e^c ), \eqno\stepeq
$$
where $a$, $b$, $c$ are space indices.  The mutual friction forces
are given in equations (14), (15), and (19) in Mendell (1991b).  Keeping
only the largest terms in those equations, and making the same
approximations as discussed at the beginning of this section,
one can show that
$$
\delta \bmath{F}_{n({\rm mf})} = - {\rm B}_n {\varrho^2 \over \rhon \rhop}
\vert {\bomega}_o \vert \left \{ \delta \bmath{V}
- {{\bomega}_o ({\bomega}_o \cdot \delta \bmath{V}) 
\over \vert {\bomega}_o \vert^2} \right \} , \eqno\stepeq
$$
$$
\delta \bmath{F}_{p({\rm mf})} = - {\rm B}_p 
\left ( {\varepsilon_p \over \rhop \Phi_o} \right )
({\bnabla} \times \delta \bmath{B}) , \eqno\stepeq
$$
where $\delta \bmath{V} =  \delta \bmath{v}_n - \delta \bmath{u}$.
A term proportional to $\bmath{B}_o \cdot ({\bnabla} \times \delta \bmath{B})$ has been
dropped in equation (92) in keeping with the approximate form for
$\delta {\blambda}_p$ given in equation (85).  Simple estimates for the
mutual friction coefficients based on the formulas given
in Mendell (1991b) are
$$
{\rm B}_n \approx .01 (\rhonp/\rhopp)^2
(\rhopp/\rhop)^{1/2}(\rhop/\rhon) \approx 10^{-4}, \eqno\stepeq
$$
$$
{\rm B}_p \approx .01 (\rhoe / \rhop) \approx 10^{-5}. \eqno\stepeq
$$

It will now be possible to find an energy functional from which the damping time
$\tau$ of modes with time dependence given by ${\rm exp}(i\omega t - t/\tau)$
can more easily be found.  Generalizing equation (32)
given in Mendell (1991b) an appropriate real energy functional is
$$
\eqalign {
& E = {1 \over 2} \int \Biggl \{ \rhonn \delta \bmath{v}_n \cdot \delta \bmath{v}_{n}^*
+ \rhonp \delta \bmath{v}_n \cdot \delta \bmath{u}^* \cr
& \, + \rhonp \delta \bmath{u} \cdot \delta \bmath{v}_n^*
+ \rhopp \delta \bmath{u} \cdot \delta \bmath{u}^*
+ \sum_{AB} {\partial^2 U_I \over
\partial \rho_A \partial \rho_B} \delta \rho_A \delta \rho_B^* \cr
& \, - {{\bnabla} \delta \varphi \cdot {\bnabla} \delta \varphi^* \over 4 \pi G}
+ {\varepsilon_p \delta \bmath{B} \cdot \delta \bmath{B}^*
 \over \Phi_o \vert \bmath{B}_o \vert} \Biggr \} d^3 x . \cr } \eqno\stepeq
$$
In this equation $A$ and $B$ are indices which take on the values $(n, p, e)$
and $U_I$ is the internal energy of the fluid
($\partial U_I / \partial \rho_A = \mu_A$).
The damping time $\tau$ of a mode can then be calculated using
$$
{1 \over \tau} = - {1 \over 2E}{dE \over dt} . \eqno\stepeq
$$
Using the equations in Section 2 and the approximations of this
section to compute $dE/dt$, and writing the result in the form
$1/\tau = 1/\tau_\eta + 1/\tau_{{\rm mf}}$,
the shear damping time and the mutual friction damping time can
be shown to be
$$
{1 \over \tau_\eta} = {1 \over 2E} \int 2\eta_e \delta \Theta^{ab}
\delta \Theta_{ab}^* d^3 x , \eqno\stepeq
$$
$$
\eqalign{
{ 1 \over \tau_{{\rm mf}} } = {1 \over 2E} \int \Biggl \{ \, &
{\rm B}_n \rhon { \vert {\bomega}_o \vert
\varrho^4 \over \rhon^2\rhop^2}
\left [ \vert \delta \bmath{V} \vert^2
-  { \vert {\bomega}_o \cdot \delta \bmath{V} \vert^2
\over \vert {\bomega}_o \vert^2 } \right ] \cr
& {\rm B}_p { \varepsilon_p^2 \over \rhop \Phi_o^2 a_p \vert \bmath{B}_o \vert }
\vert {\bnabla} \times \delta \bmath{B} \vert^2
\Biggr \} d^3 x . \cr
} \eqno\stepeq
$$

The damping times of the modes found in Section 3 and Section 4 can now be estimated.  First,
equations (80), (36), and (76) are used to relate $\delta \bmath{v}_n$,
$\delta \bmath{v}_e$, and $\delta \bmath{B}$ to $\delta \bmath{u}$, and the results
then are substituted into equations (97) and (98).  It will
be convenient to express the damping times in terms of the quality of the
mode, $Q = \omega \tau$.  For the cyclotron-vortex frequency, given in equation
(59), the viscous damping time dominates, and
$ Q_{{\rm {\small CV}}} = \omega_{{\rm {\small CV}}} (\rhop / \rhopp)
[\rhop /(\eta_e k^2)]$, or
$$
Q_{{\rm {\small CV}}} 
\approx 2.7 \times 10^3
\lambda_{\ssix}
T_{e\sseven}{}^2
\rho_{\sfifteen}{}^{-1}
\left ( {\rhop \over \rho} \right ) \,
\sqrt {(1 + \Delta) B_{o\stwelve}}
\, .  \eqno\stepeq
$$
For the rotating case, the frequency splits into the three branches given in
equation (84), and viscosity dominates the top branch, while
mutual friction due to electron scattering off the neutron vortices
dominates the damping of the middle and bottom branches.  The resulting $Q$
for each of these modes is
$$
Q = \left \{ \eqalign {
& {\Omega \rho \over \eta_e k^2} \approx 10^4
\Omega_{\stwo}
\rho_{\sfifteen}{}^{-1}
\lambda_{\ssix}{}^2
T_{e\sseven}{}^2 \cr
& [{\rm B}_n (\rhon / \rhop)]^{-1} \approx 100
(\rhopp / \rhonp)^2
(\rhop / \rhopp)^{1/2} \cr
& [{\rm B}_n (\rhon / \rhop)]^{-1} \approx 100
(\rhopp / \rhonp)^2
(\rhop / \rhopp)^{1/2} \cr
 } \right. \eqno\stepeq
$$

\section{Magnetic Spin-up of the Core}

It has been suggested that magnetic fields are responsible for the
spin-up of the neutron star core during a pulsar glitch.  Detailed studies
of the spin-up process are given in Easson (1979), Chao, Cheng \& Ding (1992),
Sedrakian, et al. (1995) and Abney \& Epstein (1996).
It also has been suggested that Alfv\'{e}n waves might be excited during
the glitch (cf., Baym et al. 1969; Van Horn 1980 and references therein.)
The goal of this section is to study the influence of cyclotron-vortex waves
on the crust-core coupling time, and to place limits on the interior magnetic
field.  First the simple method of Abney et al. (1996) is used, followed by
estimates based on the method of Easson (1979).  The results are compared
with the other studies mentioned above.

Recently Abney et al. (1996) have shown that the rapid
spin-up time for the 1988 `Christmas' glitch in the Vela pulsar
implies a crust-core coupling time of less than $10$ seconds.
They also argue, following Abney \& Epstein (1996), that Ekman pumping
and viscosity cannot be responsible for this short timescale.  
Following their `back of the envelope' derivation, any wave with
phase velocity $v = \omega /k $ associated with the forces that
spin up the core must obey the approximate constraint
$$
{R \over v} = {R k \over \omega} \le t_{\rm {\small CI}} , \eqno\stepeq
$$
where $t_{\rm {\small CI}}$ is the crust-interior coupling
time and $R$ the radius of the core. This equation can be used
to derive a lower limit for the interior magnetic field.  

For an ordinary-fluid neutron star, substituting the Alfv\'{e}n wave frequency
given in equation (42) into equation (101) and solving for $B_o$ gives a lower
bound on the interior magnetic field of
$$
B_o \ge 10^{13}
\rho_{\sfifteen}{}^{1/2}
R_{\ssix}
t_{{\rm {\small CI}}\sten}{}^{-1}
\; {\rm G} , \eqno\stepeq
$$
where $R_{\ssix}$ is $R$ divided by $10^6 \; {\rm cm}$ and
$t_{{\rm {\small CI}}\sten}$ is $t_{{\rm {\small CI}}}$ divided by $10 \; {\rm s}$.
This is exactly the same lower bound that Abney et al. (1996) give for an
ordinary-fluid neutron star.

However it has been shown here that Alfv\'{e}n waves do not exist
in superconducting-superfluid neutron stars, but are instead replaced
by cyclotron-vortex waves.  Substituting equation (59)
into equation (101) and solving for $B_o$ gives
$$
B_o \ge {10^{10} \over 1 + \Delta} 
\left ( {\rhop \over \rhopp} \right )^2
R_{\ssix}{}^2
t_{{\rm {\small CI}}\sten}{}^{-2}
\; {\rm G} . \eqno\stepeq
$$
To compare with Abney et al. (1996), in this case they 
assume that confining the flux to the proton vortices increases
the Alfv\'{e}n phase velocity by a factor of $\sqrt{H_{c1}/B_o}$
[equals the square root of the vortex energy density over the
magnetic energy density, see eq. (60) and Easson \& Pethick 1977]
and use $H_{c1} \approx 10^{15} {\rm G}$.  This gives roughly the cyclotron-vortex
wave velocity. Thus their result is fundamentally the same as that given here,
though they have ignored the mass density dependance
of $H_{c1}$ and do not find the factor $(\rhop/\rhopp)^2$.

To improve on the simple method of Abney et al. (1996) one should 
account for the effects of rotation, compressibility, stratification,
dissipation, gravity, and geometry on the spin-up.  Previous studies have
shown that the spin-up of a rotating fluid proceeds by boundary-layer formation,
followed by suction into the boundary layer (e.g., Ekman pumping,
as mentioned above), and finally damping of any residual oscillations
(Easson 1979; Abney \& Epstein 1996; and references therein).
Easson (1979) considered the spin-up of a constant-density
rotating-slab model using MHD equations for an ordinary p-e plasma,
and concluded that suction into a boundary layer was the most
important process, even in the case of magnetic spin-up.
The result given in that paper for the magnetic spin-up time of the core
depends on both the angular velocity and the Alfv\'{e}n frequency, and is given
by $({\Omega k R / \omega_{\rm {\small A}}})^{2/3} (2 / \Omega)$.
However, Abney \& Epstein (1996) have shown that stratification (and compressibility)
in a neutron star can limit the spin-up during the suction phase to only a
very small region near the crust-core boundary, greatly increasing the spin-up
time of the rest of the interior fluid.  That paper only considered the effects of
viscosity on the spin-up, but its general conclusions should apply in the magnetic
case as well.  Easson's magnetic spin-up time then would only apply to the
crust-core boundary-layer, which he found has an approximate thickness of
$\omega_{\rm {\small A}}/(\Omega k)$ (the distance an Alfv\'{e}n wave
penetrates the core during one rotation).  For the entire core to spin-up, one
can then imagine a succession of boundary layers forming from the crust
down to the centre of the star.  Each layer would
spin up on a time scale approximately equal to Easson's magnetic spin-up time
and the number of such layers would be ${\Omega k R / \omega_{\rm {\small A}}}$.
This gives a total magnetic spin-up time of 
$({\Omega k R / \omega_{\rm {\small A}}})^{5/3} (2 / \Omega)$.  Obviously this
represents an upper bound on the spin-up time, and Easson's result represents
a lower bound.  To summarize, if magnetic fields are responsible for the core
spin-up, one can tentatively predict that for an ordinary-fluid
neutron star the crust-core coupling time will be constrained by, 
$$
\left ( {\Omega k R \over
\omega_{\rm {\small A}}} \right )^{2/3} {2 \over \Omega}
\le t_{\rm {\small CI}} \le
\left ( {\Omega k R \over
\omega_{\rm {\small A}}} \right )^{5/3} {2 \over \Omega}. \eqno\stepeq
$$
Ignoring dissipative effects and the dynamics of the neutrons, 
it can be shown that the simplified MHD equations given
at the beginning of Section 5 are identical to those in Easson
(1979) but with the magnetic force in that paper increased by
$(\omega_{\rm {\small CV}} / \omega_{\rm {\small A}})^2$
and the velocity field in that paper replaced
with $(\rhopp / \rhop) \bmath{u}$.  Thus the corresponding limits
on $t_{\rm {\small CI}}$ in the superconducting-superfluid case are
identical to those in equation (104), but with the Alfv\'{e}n frequency
replaced with the cyclotron-vortex frequency, i.e.,
$$
\left ( {\Omega k R \over
\omega_{\rm {\small CV}}} \right )^{2/3} {2 \over \Omega}
\le t_{\rm {\small CI}} \le
\left ( {\Omega k R \over
\omega_{\rm {\small CV}}} \right )^{5/3} {2 \over \Omega}. \eqno\stepeq
$$
Easson correctly predicts the same lower limit given here
(up to factors of $\rhopp/\rhop$) by assuming that the
Alfv\'{e}n phase velocity is increased by a factor of
$\sqrt{H_{c1}/B_o}$ (the same factor used by Abney et al.
1996 as described above). 

Using equation (42) for the Alfv\'{e}n frequency
and equation (59) for the cyclotron-vortex frequency
in equations (104) and (105) gives the following
limits on the interior magnetic field:
$$
10^{12}
{\rho_{\sfifteen}{}^{1/2}
R_{\ssix}
\over
\Omega_{\stwo}{}^{1/2}
t_{{\rm {\small CI}}\sten}{}^{3/2}}
\; {\rm G}
\le B_o \le
10^{14}
{\rho_{\sfifteen}{}^{1/2}
R_{\ssix}
\Omega_{\stwo}{}^{2/5}
\over 
t_{{\rm {\small CI}}\sten}{}^{3/5}}
\; {\rm G} , \eqno\stepeq
$$
in the ordinary fluid case, and 
$$
{10^{8} \over 1 + \Delta}
{\rhop^2 \over \rhopp^2}
{R_{\ssix}^2
\over
\Omega_{\stwo}
t_{{\rm {\small CI}}\sten}{}^{3}}
\; {\rm G}
\le B_o \le
{10^{13} \over 1 + \Delta}
{\rhop^2 \over \rhopp^2}
{R_{\ssix}^2
\Omega_{\stwo}{}^{4/5}
\over
t_{{\rm {\small CI}}\sten}{}^{6/5}}
\; {\rm G} , \eqno\stepeq
$$
in the superconducting-superfluid case.
Note that the lower bounds in these equations are
much less stringent that those given by Abney et al.
(1996), and also in equations (102) and (103) (e.g., consider
the Vela pulsar with $\Omega = 70.6 \; {\rm s}^{-1}$ and 
$t_{{\rm {\small CI}}}$ close to $10 \; {\rm s}$).
But if stratification is as important in the case of magnetic spin-up
as the study of Abney \& Epstein (1996) implies, then the lower bounds of
Abney et al. are likely to be correct.  A more rigorous study
of the spin-up process is needed to confirm the upper bounds
given above, and may in fact show that the lower bounds should
actually be very close to these upper bounds.

Without direct observation it is hard to say much with certainty
about the interior magnetic field of a neutron star.  An early
estimate predicted it would persist for the lifetime
of the universe (Baym, Pethick \& Pines 1969), but other studies
suggest that it might decay in less than $10^7 {\rm yrs}$
(Jones 1987; Haensel, Urpin \& Yakovlev 1990) or that pinning between neutron
and proton vortices causes a portion of the interior flux to be expelled
during spin down (Srinivasan et al. 1990; Jones 1991; Ding, Cheng \& Chau 1993).
These latter models of flux expulsion give results that are consistent with
the lower bound given in equation (107), but are harder to reconcile with
the lower bounds given in the other equations.

In contrast to the lower bounds given here,  Chao, Cheng \& Ding (1992)
actually find an upper bound on the interior magnetic field of approximately
$B_o \le 10^{8-9} {\rm G}$ using vortex creep theory to study the core spin-up.
They assume that the neutron vortices are pinned by their magnetic interaction
with the proton vortices, but as they point out [and see eq. (17)] other
forces dominate the pinning forces during the initial spin-up.  It is possible that
pinning and vortex creep could be important during the transition from spin-up
to post-glitch relaxation back to equilibrium.  In even further contrast to
the results given here, Sedrakian et al.(1995) use a vortex cluster model
which predicts that proton vortices cluster around the magnetized
neutron vortices.  This produces an interior magnetic field that is tied to
the star's rotation rate and in their model the spin-up is independent of
any relic field present at the star's birth.

It is hoped future observations of pulsar glitches that time-resolve the spin-up
will help distinguish between the various models mention here.

\section{Discussion}

The MHD equations for the outer-core of a superconducting-superfluid neutron star have
been presented in equations (24)--(34), while a greatly simplified form of
the equations for the vector fields is given in equations (86)--(88).
It has been shown that Alfv\'{e}n waves are replaced by cyclotron-vortex waves in this
region, and it has been shown how these waves and the `drag' effect modify
the inertial mode when rotation is included [see eqs. (59) and (84)].  
The dissipative time scales due to electron viscosity and mutual friction can
be computed using equations (95)--(98), and the damping times of the modes
discussed in this paper can be found in equations (99)--(100).  The 
magnetic spin-up of the core during a pulsar glitch is discussed in Section 6, and
various limits on the interior magnetic field are found and compared with other
results.

It is intriging to think that glitch observations can constrain the physics
of the neutron-star core.  The limits given in Section 6 should be improved upon
by solving the MHD equations given in this paper for the core spin-up,
either by using the analytic methods of Easson (1979) or Abney \& Epstein (1996),
or numerically.  The effects of the magnetic field, rotation,
compressibility, stratification, dissipation, gravity, and geometry
on the spin-up could then be studied.  Other areas that require further
investigation are the crust-core boundary conditions (including surface
pinning effects), the effects of having an inner core, and the
relativistic corrections to the equations.  The results of future studies
could then be compared with the competing models mentioned in Section 6, and
ideally it will also be possible to make a comparison with future observations.  

\section*{Acknowledgments}

I wish to thank Steve Detweiler and Jim Ipser for reading a rough draft of this
manuscript, Richard Epstein and Angela Olinto for discussing Alfv\'{e}n waves
in neutron stars, and Richard Epstein for discussing the core spin-up process.
This research has been supported by the Institute for Fundamental Theory at the
University of Florida.

\bye

%% file: mn.tex
%
%
%
%

\catcode `\@=11 

\def\@version{1.6}
\def\@verdate{18th September 1995}

%
%


\newif\ifprod@font

\ifx\@typeface\undefined
  \def\@typeface{Comp. Modern}\prod@fontfalse
\else
  \prod@fonttrue 
\fi

\def\newfam{\alloc@8\fam\chardef\sixt@@n} 

\ifprod@font
\font\fiverm=mtr10 at 5pt
\font\fivebf=mtbx10 at 5pt
\font\fiveit=mtti10 at 5pt
\font\fivesl=mtsl10 at 5pt
\font\fivett=cmtt8 at 5pt     \hyphenchar\fivett=-1
\font\fivecsc=mtcsc10 at 5pt
\font\fivesf=mtss10 at 5pt
\font\fivei=mtmi10 at 5pt      \skewchar\fivei='177
\font\fivesy=mtsy10 at 5pt     \skewchar\fivesy='60

\font\sixrm=mtr10 at 6pt
\font\sixbf=mtbx10 at 6pt
\font\sixit=mtti10 at 6pt
\font\sixsl=mtsl10 at 6pt
\font\sixtt=cmtt8 at 6pt      \hyphenchar\sixtt=-1
\font\sixcsc=mtcsc10 at 6pt
\font\sixsf=mtss10 at 6pt
\font\sixi=mtmi10 at 6pt       \skewchar\sixi='177
\font\sixsy=mtsy10 at 6pt      \skewchar\sixsy='60

\font\sevenrm=mtr10 at 7pt
\font\sevenbf=mtbx10 at 7pt
\font\sevenit=mtti10 at 7pt
\font\sevensl=mtsl10 at 7pt
\font\seventt=cmtt8 at 7pt     \hyphenchar\seventt=-1
\font\sevencsc=mtcsc10 at 7pt
\font\sevensf=mtss10 at 7pt
\font\seveni=mtmi10 at 7pt      \skewchar\seveni='177
\font\sevensy=mtsy10 at 7pt     \skewchar\sevensy='60

\font\eightrm=mtr10 at 8pt
\font\eightbf=mtbx10 at 8pt
\font\eightit=mtti10 at 8pt
\font\eighti=mtmi10 at 8pt      \skewchar\eighti='177
\font\eightsy=mtsy10 at 8pt     \skewchar\eightsy='60
\font\eightsl=mtsl10 at 8pt
\font\eighttt=cmtt8             \hyphenchar\eighttt=-1
\font\eightcsc=mtcsc10 at 8pt
\font\eightsf=mtss10 at 8pt

\font\ninerm=mtr10 at 9pt
\font\ninebf=mtbx10 at 9pt
\font\nineit=mtti10 at 9pt
\font\ninei=mtmi10 at 9pt      \skewchar\ninei='177
\font\ninesy=mtsy10 at 9pt     \skewchar\ninesy='60
\font\ninesl=mtsl10 at 9pt
\font\ninett=cmtt9             \hyphenchar\ninett=-1
\font\ninecsc=mtcsc10 at 9pt
\font\ninesf=mtss10 at 9pt

\font\tenrm=mtr10
\font\tenbf=mtbx10
\font\tenit=mtti10
\font\teni=mtmi10		\skewchar\teni='177
\font\tensy=mtsy10		\skewchar\tensy='60
\font\tenex=cmex10
\font\tensl=mtsl10
\font\tentt=cmtt10		\hyphenchar\tentt=-1
\font\tencsc=mtcsc10
\font\tensf=mtss10

\font\elevenrm=mtr10 at 11pt
\font\elevenbf=mtbx10 at 11pt
\font\elevenit=mtti10 at 11pt
\font\eleveni=mtmi10 at 11pt      \skewchar\eleveni='177
\font\elevensy=mtsy10 at 11pt     \skewchar\elevensy='60
\font\elevensl=mtsl10 at 11pt
\font\eleventt=cmtt10 at 11pt     \hyphenchar\eleventt=-1
\font\elevencsc=mtcsc10 at 11pt
\font\elevensf=mtss10 at 11pt

\font\twelverm=mtr10 at 12pt
\font\twelvebf=mtbx10 at 12pt
\font\twelveit=mtti10 at 12pt
\font\twelvesl=mtsl10 at 12pt
\font\twelvett=cmtt12             \hyphenchar\twelvett=-1
\font\twelvecsc=mtcsc10 at 12pt
\font\twelvesf=mtss10 at 12pt
\font\twelvei=mtmi10 at 12pt      \skewchar\twelvei='177
\font\twelvesy=mtsy10 at 12pt     \skewchar\twelvesy='60

\font\fourteenrm=mtr10 at 14pt
\font\fourteenbf=mtbx10 at 14pt
\font\fourteenit=mtti10 at 14pt
\font\fourteeni=mtmi10 at 14pt      \skewchar\fourteeni='177
\font\fourteensy=mtsy10 at 14pt     \skewchar\fourteensy='60
\font\fourteensl=mtsl10 at 14pt
\font\fourteentt=cmtt12 at 14pt     \hyphenchar\fourteentt=-1
\font\fourteencsc=mtcsc10 at 14pt
\font\fourteensf=mtss10 at 14pt

\font\seventeenrm=mtr10 at 17pt
\font\seventeenbf=mtbx10 at 17pt
\font\seventeenit=mtti10 at 17pt
\font\seventeeni=mtmi10 at 17pt      \skewchar\seventeeni='177
\font\seventeensy=mtsy10 at 17pt     \skewchar\seventeensy='60
\font\seventeensl=mtsl10 at 17pt
\font\seventeentt=cmtt12 at 17pt     \hyphenchar\seventeentt=-1
\font\seventeencsc=mtcsc10 at 17pt
\font\seventeensf=mtss10 at 17pt
\else
\font\fiverm=cmr5
\font\fivei=cmmi5             \skewchar\fivei='177
\font\fivesy=cmsy5            \skewchar\fivesy='60
\font\fivebf=cmbx5

\font\sixrm=cmr6
\font\sixi=cmmi6             \skewchar\sixi='177
\font\sixsy=cmsy6            \skewchar\sixsy='60
\font\sixbf=cmbx6

\font\sevenrm=cmr7
\font\sevenit=cmti7
\font\seveni=cmmi7             \skewchar\seveni='177
\font\sevensy=cmsy7            \skewchar\sevensy='60
\font\sevenbf=cmbx7

\font\eightrm=cmr8
\font\eightbf=cmbx8
\font\eightit=cmti8
\font\eighti=cmmi8			\skewchar\eighti='177
\font\eightsy=cmsy8			\skewchar\eightsy='60
\font\eightsl=cmsl8
\font\eighttt=cmtt8			\hyphenchar\eighttt=-1
\font\eightcsc=cmcsc10 at 8pt
\font\eightsf=cmss8

\font\ninerm=cmr9
\font\ninebf=cmbx9
\font\nineit=cmti9
\font\ninei=cmmi9			\skewchar\ninei='177
\font\ninesy=cmsy9			\skewchar\ninesy='60
\font\ninesl=cmsl9
\font\ninett=cmtt9			\hyphenchar\ninett=-1
\font\ninecsc=cmcsc10 at 9pt
\font\ninesf=cmss9

\font\tenrm=cmr10
\font\tenbf=cmbx10
\font\tenit=cmti10
\font\teni=cmmi10		\skewchar\teni='177
\font\tensy=cmsy10		\skewchar\tensy='60
\font\tenex=cmex10
\font\tensl=cmsl10
\font\tentt=cmtt10		\hyphenchar\tentt=-1
\font\tencsc=cmcsc10
\font\tensf=cmss10

\font\elevenrm=cmr10 scaled \magstephalf
\font\elevenbf=cmbx10 scaled \magstephalf
\font\elevenit=cmti10 scaled \magstephalf
\font\eleveni=cmmi10 scaled \magstephalf	\skewchar\eleveni='177
\font\elevensy=cmsy10 scaled \magstephalf	\skewchar\elevensy='60
\font\elevensl=cmsl10 scaled \magstephalf
\font\eleventt=cmtt10 scaled \magstephalf	\hyphenchar\eleventt=-1
\font\elevencsc=cmcsc10 scaled \magstephalf
\font\elevensf=cmss10 scaled \magstephalf

\font\twelverm=cmr10 scaled \magstep1
\font\twelvebf=cmbx10 scaled \magstep1
\font\twelvei=cmmi10 scaled \magstep1      \skewchar\twelvei='177
\font\twelvesy=cmsy10 scaled \magstep1     \skewchar\twelvesy='60

\font\fourteenrm=cmr10 scaled \magstep2
\font\fourteenbf=cmbx10 scaled \magstep2
\font\fourteenit=cmti10 scaled \magstep2
\font\fourteeni=cmmi10 scaled \magstep2		\skewchar\fourteeni='177
\font\fourteensy=cmsy10 scaled \magstep2	\skewchar\fourteensy='60
\font\fourteensl=cmsl10 scaled \magstep2
\font\fourteentt=cmtt10 scaled \magstep2	\hyphenchar\fourteentt=-1
\font\fourteencsc=cmcsc10 scaled \magstep2
\font\fourteensf=cmss10 scaled \magstep2

\font\seventeenrm=cmr10 scaled \magstep3
\font\seventeenbf=cmbx10 scaled \magstep3
\font\seventeenit=cmti10 scaled \magstep3
\font\seventeeni=cmmi10 scaled \magstep3	\skewchar\seventeeni='177
\font\seventeensy=cmsy10 scaled \magstep3	\skewchar\seventeensy='60
\font\seventeensl=cmsl10 scaled \magstep3
\font\seventeentt=cmtt10 scaled \magstep3	\hyphenchar\seventeentt=-1
\font\seventeencsc=cmcsc10 scaled \magstep3
\font\seventeensf=cmss10 scaled \magstep3
\fi

\def\hexnumber#1{\ifcase#1 0\or1\or2\or3\or4\or5\or6\or7\or8\or9\or
  A\or B\or C\or D\or E\or F\fi}

\def\makestrut{%
  \setbox\strutbox=\hbox{%
    \vrule height.7\baselineskip depth.3\baselineskip width \z@}%
}

\def\baselinestretch{1}
\newskip\tmp@bls

\def\b@ls#1{
  \tmp@bls=#1\relax
  \baselineskip=#1\relax\makestrut
  \normalbaselineskip=\baselinestretch\tmp@bls
  \normalbaselines
}

\def\nostb@ls#1{
  \normalbaselineskip=#1\relax
  \normalbaselines
  \makestrut
}

%

\newfam\scfam  
\newfam\sffam  

\def\mit{\fam\@ne}
\def\cal{\fam\tw@}
\def\em{\ifdim\fontdimen1\font>\z@ \rm\else\it\fi}

\textfont3=\tenex
\scriptfont3=\tenex
\scriptscriptfont3=\tenex

\setbox0=\hbox{\tenex B} \p@renwd=\wd0 

\def\eightpoint{
  \def\rm{\fam0\eightrm}%
  \textfont0=\eightrm \scriptfont0=\sixrm \scriptscriptfont0=\fiverm%
  \textfont1=\eighti  \scriptfont1=\sixi  \scriptscriptfont1=\fivei%
  \textfont2=\eightsy \scriptfont2=\sixsy \scriptscriptfont2=\fivesy%
  \textfont\itfam=\eightit\def\it{\fam\itfam\eightit}%
  \ifprod@font
    \scriptfont\itfam=\sixit
      \scriptscriptfont\itfam=\fiveit
  \else
    \scriptfont\itfam=\eightit
      \scriptscriptfont\itfam=\eightit
  \fi
  \textfont\bffam=\eightbf%
    \scriptfont\bffam=\sixbf%
      \scriptscriptfont\bffam=\fivebf%
  \def\bf{\fam\bffam\eightbf}%
  \textfont\slfam=\eightsl\def\sl{\fam\slfam\eightsl}%
  \ifprod@font
    \scriptfont\slfam=\sixsl
      \scriptscriptfont\slfam=\fivesl
  \else
    \scriptfont\slfam=\eightsl
      \scriptscriptfont\slfam=\eightsl
  \fi
  \textfont\ttfam=\eighttt\def\tt{\fam\ttfam\eighttt}%
  \ifprod@font
    \scriptfont\ttfam=\sixtt
      \scriptscriptfont\ttfam=\fivett
  \else
    \scriptfont\ttfam=\eighttt
      \scriptscriptfont\ttfam=\eighttt
  \fi
  \textfont\scfam=\eightcsc\def\sc{\fam\scfam\eightcsc}%
  \ifprod@font
    \scriptfont\scfam=\sixcsc
      \scriptscriptfont\scfam=\fivecsc
  \else
    \scriptfont\scfam=\eightcsc
      \scriptscriptfont\scfam=\eightcsc
  \fi
  \textfont\sffam=\eightsf\def\sf{\fam\sffam\eightsf}%
  \ifprod@font
    \scriptfont\sffam=\sixsf
      \scriptscriptfont\sffam=\fivesf
  \else
    \scriptfont\sffam=\eightsf
      \scriptscriptfont\sffam=\eightsf
  \fi
  \def\oldstyle{\fam\@ne\eighti}%
  \b@ls{10pt}\rm\@viiipt%
}
\def\@viiipt{}

\def\ninepoint{
  \def\rm{\fam0\ninerm}%
  \textfont0=\ninerm \scriptfont0=\sixrm \scriptscriptfont0=\fiverm%
  \textfont1=\ninei  \scriptfont1=\sixi  \scriptscriptfont1=\fivei%
  \textfont2=\ninesy \scriptfont2=\sixsy \scriptscriptfont2=\fivesy%
  \textfont\itfam=\nineit\def\it{\fam\itfam\nineit}%
  \ifprod@font
    \scriptfont\itfam=\sixit
      \scriptscriptfont\itfam=\fiveit
  \else
    \scriptfont\itfam=\nineit
      \scriptscriptfont\itfam=\nineit
  \fi
  \textfont\bffam=\ninebf%
    \scriptfont\bffam=\sixbf%
      \scriptscriptfont\bffam=\fivebf%
  \def\bf{\fam\bffam\ninebf}%
  \textfont\slfam=\ninesl\def\sl{\fam\slfam\ninesl}%
  \ifprod@font
    \scriptfont\slfam=\sixsl
      \scriptscriptfont\slfam=\fivesl
  \else
    \scriptfont\slfam=\ninesl
      \scriptscriptfont\slfam=\ninesl
  \fi
  \textfont\ttfam=\ninett\def\tt{\fam\ttfam\ninett}%
  \ifprod@font
    \scriptfont\ttfam=\sixtt
      \scriptscriptfont\ttfam=\fivett
  \else
    \scriptfont\ttfam=\ninett
      \scriptscriptfont\ttfam=\ninett
  \fi
  \textfont\scfam=\ninecsc\def\sc{\fam\scfam\ninecsc}%
  \ifprod@font
    \scriptfont\scfam=\sixcsc
      \scriptscriptfont\scfam=\fivecsc
  \else
    \scriptfont\scfam=\ninecsc
      \scriptscriptfont\scfam=\ninecsc
  \fi
  \textfont\sffam=\ninesf\def\sf{\fam\sffam\ninesf}%
  \ifprod@font
    \scriptfont\sffam=\sixsf
      \scriptscriptfont\sffam=\fivesf
  \else
    \scriptfont\sffam=\ninesf
      \scriptscriptfont\sffam=\ninesf
  \fi
  \def\oldstyle{\fam\@ne\ninei}%
  \b@ls{\TextLeading plus \Feathering}\rm\@ixpt%
}
\def\@ixpt{}

\def\tenpoint{
  \def\rm{\fam0\tenrm}%
  \textfont0=\tenrm \scriptfont0=\sevenrm \scriptscriptfont0=\fiverm%
  \textfont1=\teni  \scriptfont1=\seveni  \scriptscriptfont1=\fivei%
  \textfont2=\tensy \scriptfont2=\sevensy \scriptscriptfont2=\fivesy%
  \textfont\itfam=\tenit\def\it{\fam\itfam\tenit}%
  \ifprod@font
    \scriptfont\itfam=\sevenit
      \scriptscriptfont\itfam=\fiveit
  \else
    \scriptfont\itfam=\tenit
      \scriptscriptfont\itfam=\tenit
  \fi
  \textfont\bffam=\tenbf%
    \scriptfont\bffam=\sevenbf%
      \scriptscriptfont\bffam=\fivebf%
  \def\bf{\fam\bffam\tenbf}%
  \textfont\slfam=\tensl\def\sl{\fam\slfam\tensl}%
  \ifprod@font
    \scriptfont\slfam=\sevensl
      \scriptscriptfont\slfam=\fivesl
  \else
    \scriptfont\slfam=\tensl
      \scriptscriptfont\slfam=\tensl
  \fi
  \textfont\ttfam=\tentt\def\tt{\fam\ttfam\tentt}%
  \ifprod@font
    \scriptfont\ttfam=\seventt
      \scriptscriptfont\ttfam=\fivett
  \else
    \scriptfont\ttfam=\tentt
      \scriptscriptfont\ttfam=\tentt
  \fi
  \textfont\scfam=\tencsc\def\sc{\fam\scfam\tencsc}%
  \ifprod@font
    \scriptfont\scfam=\sevencsc
      \scriptscriptfont\scfam=\fivecsc
  \else
    \scriptfont\scfam=\tencsc
      \scriptscriptfont\scfam=\tencsc
  \fi
  \textfont\sffam=\tensf\def\sf{\fam\sffam\tensf}%
  \ifprod@font
    \scriptfont\sffam=\sevensf
      \scriptscriptfont\sffam=\fivesf
  \else
    \scriptfont\sffam=\tensf
      \scriptscriptfont\sffam=\tensf
  \fi
  \def\oldstyle{\fam\@ne\teni}%
  \b@ls{11pt}\rm\@xpt%
}
\def\@xpt{}

\def\elevenpoint{
  \def\rm{\fam0\elevenrm}%
  \textfont0=\elevenrm \scriptfont0=\eightrm \scriptscriptfont0=\sixrm%
  \textfont1=\eleveni  \scriptfont1=\eighti  \scriptscriptfont1=\sixi%
  \textfont2=\elevensy \scriptfont2=\eightsy \scriptscriptfont2=\sixsy%
  \textfont\itfam=\elevenit\def\it{\fam\itfam\elevenit}%
  \ifprod@font
    \scriptfont\itfam=\eightit
      \scriptscriptfont\itfam=\sixit
  \else
    \scriptfont\itfam=\elevenit
      \scriptscriptfont\itfam=\elevenit
  \fi
  \textfont\bffam=\elevenbf%
    \scriptfont\bffam=\eightbf%
      \scriptscriptfont\bffam=\sixbf%
  \def\bf{\fam\bffam\elevenbf}%
  \textfont\slfam=\elevensl\def\sl{\fam\slfam\elevensl}%
  \ifprod@font
    \scriptfont\slfam=\eightsl
      \scriptscriptfont\slfam=\sixsl
  \else
    \scriptfont\slfam=\elevensl
      \scriptscriptfont\slfam=\elevensl
  \fi
  \textfont\ttfam=\eleventt\def\tt{\fam\ttfam\eleventt}%
  \ifprod@font
    \scriptfont\ttfam=\eighttt
      \scriptscriptfont\ttfam=\sixtt
  \else
    \scriptfont\ttfam=\eleventt
      \scriptscriptfont\ttfam=\eleventt
  \fi
  \textfont\scfam=\elevencsc\def\sc{\fam\scfam\elevencsc}%
  \ifprod@font
    \scriptfont\scfam=\eightcsc
      \scriptscriptfont\scfam=\sixcsc
  \else
    \scriptfont\scfam=\elevencsc
      \scriptscriptfont\scfam=\elevencsc
  \fi
  \textfont\sffam=\elevensf\def\sf{\fam\sffam\elevensf}%
  \ifprod@font
    \scriptfont\sffam=\eightsf
      \scriptscriptfont\sffam=\sixsf
  \else
    \scriptfont\sffam=\elevensf
      \scriptscriptfont\sffam=\elevensf
  \fi
  \def\oldstyle{\fam\@ne\eleveni}%
  \b@ls{13pt}\rm\@xipt%
}
\def\@xipt{}

\def\fourteenpoint{
  \def\rm{\fam0\fourteenrm}%
  \textfont0\fourteenrm  \scriptfont0\tenrm  \scriptscriptfont0\sevenrm%
  \textfont1\fourteeni   \scriptfont1\teni   \scriptscriptfont1\seveni%
  \textfont2\fourteensy  \scriptfont2\tensy  \scriptscriptfont2\sevensy%
  \textfont\itfam=\fourteenit\def\it{\fam\itfam\fourteenit}%
  \ifprod@font
    \scriptfont\itfam=\tenit
      \scriptscriptfont\itfam=\sevenit
  \else
    \scriptfont\itfam=\fourteenit
      \scriptscriptfont\itfam=\fourteenit
  \fi
  \textfont\bffam=\fourteenbf%
    \scriptfont\bffam=\tenbf%
      \scriptscriptfont\bffam=\sevenbf%
  \def\bf{\fam\bffam\fourteenbf}%
  \textfont\slfam=\fourteensl\def\sl{\fam\slfam\fourteensl}%
  \ifprod@font
    \scriptfont\slfam=\tensl
      \scriptscriptfont\slfam=\sevensl
  \else
    \scriptfont\slfam=\fourteensl
      \scriptscriptfont\slfam=\fourteensl
  \fi
  \textfont\ttfam=\fourteentt\def\tt{\fam\ttfam\fourteentt}%
  \ifprod@font
    \scriptfont\ttfam=\tentt
      \scriptscriptfont\ttfam=\seventt
  \else
    \scriptfont\ttfam=\fourteentt
      \scriptscriptfont\ttfam=\fourteentt
  \fi
  \textfont\scfam=\fourteencsc\def\sc{\fam\scfam\fourteencsc}%
  \ifprod@font
    \scriptfont\scfam=\tencsc
      \scriptscriptfont\scfam=\sevencsc
  \else
    \scriptfont\scfam=\fourteencsc
      \scriptscriptfont\scfam=\fourteencsc
  \fi
  \textfont\sffam=\fourteensf\def\sf{\fam\sffam\fourteensf}%
  \ifprod@font
    \scriptfont\sffam=\tensf
      \scriptscriptfont\sffam=\sevensf
  \else
    \scriptfont\sffam=\fourteensf
      \scriptscriptfont\sffam=\fourteensf
  \fi
  \def\oldstyle{\fam\@ne\fourteeni}%
  \b@ls{17pt}\rm\@xivpt%
}
\def\@xivpt{}

\def\seventeenpoint{
  \def\rm{\fam0\seventeenrm}%
  \textfont0\seventeenrm  \scriptfont0\twelverm  \scriptscriptfont0\tenrm%
  \textfont1\seventeeni   \scriptfont1\twelvei   \scriptscriptfont1\teni%
  \textfont2\seventeensy  \scriptfont2\twelvesy  \scriptscriptfont2\tensy%
  \textfont\itfam=\seventeenit\def\it{\fam\itfam\seventeenit}%
  \ifprod@font
    \scriptfont\itfam=\twelveit
      \scriptscriptfont\itfam=\tenit
  \else
    \scriptfont\itfam=\seventeenit
      \scriptscriptfont\itfam=\seventeenit
  \fi
  \textfont\bffam=\seventeenbf%
    \scriptfont\bffam=\twelvebf%
      \scriptscriptfont\bffam=\tenbf%
  \def\bf{\fam\bffam\seventeenbf}%
  \textfont\slfam=\seventeensl\def\sl{\fam\slfam\seventeensl}%
  \ifprod@font
    \scriptfont\slfam=\twelvesl
      \scriptscriptfont\slfam=\tensl
  \else
    \scriptfont\slfam=\seventeensl
      \scriptscriptfont\slfam=\seventeensl
  \fi
  \textfont\ttfam=\seventeentt\def\tt{\fam\ttfam\seventeentt}%
  \ifprod@font
    \scriptfont\ttfam=\twelvett
      \scriptscriptfont\ttfam=\tentt
  \else
    \scriptfont\ttfam=\seventeentt
      \scriptscriptfont\ttfam=\seventeentt
  \fi
  \textfont\scfam=\seventeencsc\def\sc{\fam\scfam\seventeencsc}%
  \ifprod@font
    \scriptfont\scfam=\twelvecsc
      \scriptscriptfont\scfam=\tencsc
  \else
    \scriptfont\scfam=\seventeencsc
      \scriptscriptfont\scfam=\seventeencsc
  \fi
  \textfont\sffam=\seventeensf\def\sf{\fam\sffam\seventeensf}%
  \ifprod@font
    \scriptfont\sffam=\twelvesf
      \scriptscriptfont\sffam=\tensf
  \else
    \scriptfont\sffam=\seventeensf
      \scriptscriptfont\sffam=\seventeensf
  \fi
  \def\oldstyle{\fam\@ne\seventeeni}%
  \b@ls{20pt}\rm\@xviipt%
}
\def\@xviipt{}

\lineskip=1pt      \normallineskip=\lineskip
\lineskiplimit=\z@ \normallineskiplimit=\lineskiplimit


\def\loadboldmathnames{%
  \def\balpha{{\bmath{\alpha}}}%
  \def\bbeta{{\bmath{\beta}}}%
  \def\bgamma{{\bmath{\gamma}}}%
  \def\bdelta{{\bmath{\delta}}}%
  \def\bepsilon{{\bmath{\epsilon}}}%
  \def\bzeta{{\bmath{\zeta}}}%
  \def\boldeta{{\bmath{\eta}}}%
  \def\btheta{{\bmath{\theta}}}%
  \def\biota{{\bmath{\iota}}}%
  \def\bkappa{{\bmath{\kappa}}}%
  \def\blambda{{\bmath{\lambda}}}%
  \def\bmu{{\bmath{\mu}}}%
  \def\bnu{{\bmath{\nu}}}%
  \def\bxi{{\bmath{\xi}}}%
  \def\bpi{{\bmath{\pi}}}%
  \def\brho{{\bmath{\rho}}}%
  \def\bsigma{{\bmath{\sigma}}}%
  \def\btau{{\bmath{\tau}}}%
  \def\bupsilon{{\bmath{\upsilon}}}%
  \def\bphi{{\bmath{\phi}}}%
  \def\bchi{{\bmath{\chi}}}%
  \def\bpsi{{\bmath{\psi}}}%
  \def\bomega{{\bmath{\omega}}}%
  \def\bvarepsilon{{\bmath{\varepsilon}}}%
  \def\bvartheta{{\bmath{\vartheta}}}%
  \def\bvarpi{{\bmath{\varpi}}}%
  \def\bvarrho{{\bmath{\varrho}}}%
  \def\bvarsigma{{\bmath{\varsigma}}}%
  \def\bvarphi{{\bmath{\varphi}}}%
  \def\baleph{{\bmath{\aleph}}}%
  \def\bimath{{\bmath{\imath}}}%
  \def\bjmath{{\bmath{\jmath}}}%
  \def\bell{{\bmath{\ell}}}%
  \def\bwp{{\bmath{\wp}}}%
  \def\bRe{{\bmath{\Re}}}%
  \def\bIm{{\bmath{\Im}}}%
  \def\bpartial{{\bmath{\partial}}}%
  \def\binfty{{\bmath{\infty}}}%
  \def\bprime{{\bmath{\prime}}}%
  \def\bemptyset{{\bmath{\emptyset}}}%
  \def\bnabla{{\bmath{\nabla}}}%
  \def\btop{{\bmath{\top}}}%
  \def\bbot{{\bmath{\bot}}}%
  \def\btriangle{{\bmath{\triangle}}}%
  \def\bforall{{\bmath{\forall}}}%
  \def\bexists{{\bmath{\exists}}}%
  \def\bneg{{\bmath{\neg}}}%
  \def\bflat{{\bmath{\flat}}}%
  \def\bnatural{{\bmath{\natural}}}%
  \def\bsharp{{\bmath{\sharp}}}%
  \def\bclubsuit{{\bmath{\clubsuit}}}%
  \def\bdiamondsuit{{\bmath{\diamondsuit}}}%
  \def\bheartsuit{{\bmath{\heartsuit}}}%
  \def\bspadesuit{{\bmath{\spadesuit}}}%
  \def\bsmallint{{\bmath{\smallint}}}%
  \def\btriangleleft{{\bmath{\triangleleft}}}%
  \def\btriangleright{{\bmath{\triangleright}}}%
  \def\bbigtriangleup{{\bmath{\bigtriangleup}}}%
  \def\bbigtriangledown{{\bmath{\bigtriangledown}}}%
  \def\bwedge{{\bmath{\wedge}}}%
  \def\bvee{{\bmath{\vee}}}%
  \def\bcap{{\bmath{\cap}}}%
  \def\bcup{{\bmath{\cup}}}%
  \def\bddagger{{\bmath{\ddagger}}}%
  \def\bdagger{{\bmath{\dagger}}}%
  \def\bsqcap{{\bmath{\sqcap}}}%
  \def\bsqcup{{\bmath{\sqcup}}}%
  \def\buplus{{\bmath{\uplus}}}%
  \def\bamalg{{\bmath{\amalg}}}%
  \def\bdiamond{{\bmath{\diamond}}}%
  \def\bbullet{{\bmath{\bullet}}}%
  \def\bwr{{\bmath{\wr}}}%
  \def\bdiv{{\bmath{\div}}}%
  \def\bodot{{\bmath{\odot}}}%
  \def\boslash{{\bmath{\oslash}}}%
  \def\botimes{{\bmath{\otimes}}}%
  \def\bominus{{\bmath{\ominus}}}%
  \def\boplus{{\bmath{\oplus}}}%
  \def\bmp{{\bmath{\mp}}}%
  \def\bpm{{\bmath{\pm}}}%
  \def\bcirc{{\bmath{\circ}}}%
  \def\bbigcirc{{\bmath{\bigcirc}}}%
  \def\bsetminus{{\bmath{\setminus}}}%
  \def\bcdot{{\bmath{\cdot}}}%
  \def\bast{{\bmath{\ast}}}%
  \def\btimes{{\bmath{\times}}}%
  \def\bstar{{\bmath{\star}}}%
  \def\bpropto{{\bmath{\propto}}}%
  \def\bsqsubseteq{{\bmath{\sqsubseteq}}}%
  \def\bsqsupseteq{{\bmath{\sqsupseteq}}}%
  \def\bparallel{{\bmath{\parallel}}}%
  \def\bmid{{\bmath{\mid}}}%
  \def\bdashv{{\bmath{\dashv}}}%
  \def\bvdash{{\bmath{\vdash}}}%
  \def\bnearrow{{\bmath{\nearrow}}}%
  \def\bsearrow{{\bmath{\searrow}}}%
  \def\bnwarrow{{\bmath{\nwarrow}}}%
  \def\bswarrow{{\bmath{\swarrow}}}%
  \def\bLeftrightarrow{{\bmath{\Leftrightarrow}}}%
  \def\bLeftarrow{{\bmath{\Leftarrow}}}%
  \def\bRightarrow{{\bmath{\Rightarrow}}}%
  \def\bleq{{\bmath{\leq}}}%
  \def\bgeq{{\bmath{\geq}}}%
  \def\bsucc{{\bmath{\succ}}}%
  \def\bprec{{\bmath{\prec}}}%
  \def\bapprox{{\bmath{\approx}}}%
  \def\bsucceq{{\bmath{\succeq}}}%
  \def\bpreceq{{\bmath{\preceq}}}%
  \def\bsupset{{\bmath{\supset}}}%
  \def\bsubset{{\bmath{\subset}}}%
  \def\bsupseteq{{\bmath{\supseteq}}}%
  \def\bsubseteq{{\bmath{\subseteq}}}%
  \def\bin{{\bmath{\in}}}%
  \def\bni{{\bmath{\ni}}}%
  \def\bgg{{\bmath{\gg}}}%
  \def\bll{{\bmath{\ll}}}%
  \def\bnot{{\bmath{\not}}}%
  \def\bleftrightarrow{{\bmath{\leftrightarrow}}}%
  \def\bleftarrow{{\bmath{\leftarrow}}}%
  \def\brightarrow{{\bmath{\rightarrow}}}%
  \def\bmapstochar{{\bmath{\mapstochar}}}%
  \def\bsim{{\bmath{\sim}}}%
  \def\bsimeq{{\bmath{\simeq}}}%
  \def\bperp{{\bmath{\perp}}}%
  \def\bequiv{{\bmath{\equiv}}}%
  \def\basymp{{\bmath{\asymp}}}%
  \def\bsmile{{\bmath{\smile}}}%
  \def\bfrown{{\bmath{\frown}}}%
  \def\bleftharpoonup{{\bmath{\leftharpoonup}}}%
  \def\bleftharpoondown{{\bmath{\leftharpoondown}}}%
  \def\brightharpoonup{{\bmath{\rightharpoonup}}}%
  \def\brightharpoondown{{\bmath{\rightharpoondown}}}%
  \def\blhook{{\bmath{\lhook}}}%
  \def\brhook{{\bmath{\rhook}}}%
  \def\bldotp{{\bmath{\ldotp}}}%
  \def\bcdotp{{\bmath{\cdotp}}}%
}

\def\,{\relax\ifmmode \mskip\thinmuskip\else \thinspace\fi}
\let\protect=\relax

\long\def\@ifundefined#1#2#3{\expandafter\ifx\csname
  #1\endcsname\relax#2\else#3\fi}




\newtoks\math@groups \math@groups={}
\def\addtom@thgroup#1#2{#1\expandafter{\the#1#2}} 



\def\addtosizeh@ok#1#2#3#4{%
  \expandafter\def\csname @#1pt\endcsname{%
    \def\s@ze{#2}\def\ss@ze{#3}\def\sss@ze{#4}\the\math@groups%
  }%
}



\let\resetsizehook=\addtosizeh@ok


\ifprod@font
  \addtosizeh@ok{viii} {8} {6}  {5}
  \addtosizeh@ok{ix}   {9} {6}  {5}
  \addtosizeh@ok{x}    {10}{7}  {5}
  \addtosizeh@ok{xi}   {11}{8}  {6}
  \addtosizeh@ok{xiv}  {14}{10} {7}
  \addtosizeh@ok{xvii} {17}{12}{10}
\else
  \addtosizeh@ok{viii} {8}     {6}     {5}
  \addtosizeh@ok{ix}   {9}     {6}     {5}
  \addtosizeh@ok{x}    {10}    {7}     {5}
  \addtosizeh@ok{xi}   {10.95} {8}     {6}
  \addtosizeh@ok{xiv}  {14.4}  {10}    {7}
  \addtosizeh@ok{xvii} {17.28} {12}    {10}
\fi

\def\get@font#1#2#3{%
  \edef\fonts@ze{\romannumeral#3}
  \edef\fontn@me{\fonts@ze#1}
  \@ifundefined{\fontn@me}%
    {
     \global\expandafter\font\csname \fontn@me\endcsname=#2 at #3pt}%
    {}%
}

\def\ass@tfont#1#2{%
  \xdef\fam@name{\csname #1\endcsname}%
  \xdef\font@name{\csname #2\endcsname}%
  \let\textfont@name\font@name
  \textfont\fam@name\textfont@name
}

\def\ass@sfont#1#2{%
  \xdef\fam@name{\csname #1\endcsname}%
  \xdef\font@name{\csname #2\endcsname}%
  \let\textfont@name\font@name
  \scriptfont\fam@name\textfont@name
}

\def\ass@ssfont#1#2{%
  \xdef\fam@name{\csname #1\endcsname}%
  \xdef\font@name{\csname #2\endcsname}%
  \let\textfont@name\font@name
  \scriptscriptfont\fam@name\textfont@name
}


\def\NewSymbolFont#1#2{%
  \expandafter\ifx\csname sym#1fam\endcsname\relax 
    \expandafter\newfam\csname sym#1fam\endcsname
    \expandafter\edef\csname sym#1fam\endcsname{\the\allocationnumber}%
    \addtom@thgroup\math@groups{%
      \get@font{#1}{#2}{\s@ze}%
      \ass@tfont{sym#1fam}{\fontn@me}%
      \get@font{#1}{#2}{\ss@ze}%
      \ass@sfont{sym#1fam}{\fontn@me}%
      \get@font{#1}{#2}{\sss@ze}%
      \ass@ssfont{sym#1fam}{\fontn@me}%
    }%
  \else
    \errmessage{Family `#1' already defined}%
  \fi
}


\def\NewMathSymbol#1#2#3#4{%
  \edef\f@mly{\expandafter\hexnumber{\csname sym#3fam\endcsname}}%
  \mathchardef#1="#2\f@mly#4\relax
}


\newif\ifd@f

\def\NewMathDelimiter#1#2#3#4#5#6{%
  \d@ftrue
  \expandafter\ifx\csname sym#3fam\endcsname\relax
    \d@ffalse \errmessage{Family `#3' is not defined}%
  \fi
  \expandafter\ifx\csname sym#5fam\endcsname\relax
    \d@ffalse \errmessage{Family `#5' is not defined}%
  \fi
  \ifd@f
    \edef\f@mly{\expandafter\hexnumber{\csname sym#3fam\endcsname}}%
    \edef\f@mlytw@{\expandafter\hexnumber{\csname sym#5fam\endcsname}}%
    \xdef#1{\delimiter"#2\f@mly #4\f@mlytw@ #6\relax}%
  \fi
}


\def\setboxz@h{\setbox\z@\hbox}
\def\wdz@{\wd\z@}
\def\boxz@{\box\z@}
\def\setbox@ne{\setbox\@ne}
\def\wd@ne{\wd\@ne}

\def\math@atom#1#2{%
   \binrel@{#1}\binrel@@{#2}}
\def\binrel@#1{\setboxz@h{\thinmuskip0mu
  \medmuskip\m@ne mu\thickmuskip\@ne mu$#1\m@th$}%
 \setbox@ne\hbox{\thinmuskip0mu\medmuskip\m@ne mu\thickmuskip
  \@ne mu${}#1{}\m@th$}%
 \setbox\tw@\hbox{\hskip\wd@ne\hskip-\wdz@}}
\def\binrel@@#1{\ifdim\wd2<\z@\mathbin{#1}\else\ifdim\wd\tw@>\z@
 \mathrel{#1}\else{#1}\fi\fi}

\def\m@thit{1}

\def\set@skchar#1{\global\expandafter\skewchar
  \csname\fontn@me\endcsname=#1\relax}

\def\NewMathAlphabet#1#2#3{%
  \def\tst{#3}%
  \ifx\tst\empty\else 
    \expandafter\gdef\csname #1@sc\endcsname{}
  \fi
  \expandafter\def\csname #1\endcsname{
    \protect\csname @#1\endcsname}%
  \expandafter\def\csname @#1\endcsname##1{
    {%
    \begingroup
      \get@font{#1}{#2}{\s@ze}%
      \@ifundefined{#1@sc}{}{\set@skchar{#3}}%
      \ass@tfont{m@thit}{\fontn@me}%
      \get@font{#1}{#2}{\ss@ze}%
      \@ifundefined{#1@sc}{}{\set@skchar{#3}}%
      \ass@sfont{m@thit}{\fontn@me}%
      \get@font{#1}{#2}{\sss@ze}%
      \@ifundefined{#1@sc}{}{\set@skchar{#3}}%
      \ass@ssfont{m@thit}{\fontn@me}%
      \math@atom{##1}{%
      \mathchoice%
        {\hbox{$\m@th\displaystyle##1$}}%
        {\hbox{$\m@th\textstyle##1$}}%
        {\hbox{$\m@th\scriptstyle##1$}}%
        {\hbox{$\m@th\scriptscriptstyle##1$}}}%
    \endgroup
    }%
  }%
}


\newif\iffirstta  \firsttatrue

\def\set@hchar#1{\global\expandafter\hyphenchar
  \csname\fontn@me\endcsname=#1\relax}

\def\NewTextAlphabet#1#2#3{%
  \iffirstta
    \global\firsttafalse
    \newfam\scratchfam
    \edef\scrt@fam{\the\allocationnumber}%
  \fi
  \def\tst{#3}%
  \ifx\tst\empty\else 
    \expandafter\gdef\csname #1@hc\endcsname{}
  \fi
  \expandafter\def\csname #1\endcsname{
    \protect\csname t@#1\endcsname}%
  \long\expandafter\def\csname t@#1\endcsname##1{
    \ifmmode
      \typeout{Warning: do not use \expandafter\string\csname #1\endcsname
        \space in math mode}\fi%
    {%
      \get@font{#1}{#2}{\s@ze}\let\t@xtfnt=\fontn@me\relax
      \@ifundefined{#1@hc}{}{\set@hchar{#3}}%
      \ass@tfont{scrt@fam}{\fontn@me}%
      \get@font{#1}{#2}{\ss@ze}%
      \@ifundefined{#1@hc}{}{\set@hchar{#3}}%
      \ass@sfont{scrt@fam}{\fontn@me}%
      \get@font{#1}{#2}{\sss@ze}%
      \@ifundefined{#1@hc}{}{\set@hchar{#3}}%
      \ass@ssfont{scrt@fam}{\fontn@me}%
      \fam\scratchfam\csname\t@xtfnt\endcsname
    ##1%
    }%
  }%
  \expandafter\def\csname #1shape
    \endcsname{\protect\csname @#1shape\endcsname}%
  \expandafter\def\csname @#1shape\endcsname{
    \ifmmode
      \typeout{Warning: do not use \expandafter\string\csname
        #1shape\endcsname \space in math mode}\fi
      \get@font{#1}{#2}{\s@ze}\let\t@xtfnt=\fontn@me\relax
      \@ifundefined{#1@hc}{}{\set@hchar{#3}}%
      \ass@tfont{scrt@fam}{\fontn@me}%
      \get@font{#1}{#2}{\ss@ze}%
      \@ifundefined{#1@hc}{}{\set@hchar{#3}}%
      \ass@sfont{scrt@fam}{\fontn@me}%
      \get@font{#1}{#2}{\sss@ze}%
      \@ifundefined{#1@hc}{}{\set@hchar{#3}}%
      \ass@ssfont{scrt@fam}{\fontn@me}%
      \fam\scratchfam\csname\t@xtfnt\endcsname
  }%
}


\ifprod@font
  \def\math@itfnt{mtmib10}
  \def\math@syfnt{mtbsy10}
\else
  \def\math@itfnt{cmmib10}
  \def\math@syfnt{cmbsy10}
\fi

\def\m@thsy{2}

\def\bmath{\protect\@bmath}
\def\@bmath#1{%
  {%
  \begingroup
    \get@font{mthit}{\math@itfnt}{\s@ze}\set@skchar{'177}%
    \ass@tfont{m@thit}{\fontn@me}%
    \get@font{mthit}{\math@itfnt}{\ss@ze}\set@skchar{'177}%
    \ass@sfont{m@thit}{\fontn@me}%
    \get@font{mthit}{\math@itfnt}{\sss@ze}\set@skchar{'177}%
    \ass@ssfont{m@thit}{\fontn@me}%
    \get@font{mthsy}{\math@syfnt}{\s@ze}\set@skchar{'60}%
    \ass@tfont{m@thsy}{\fontn@me}%
    \get@font{mthsy}{\math@syfnt}{\ss@ze}\set@skchar{'60}%
    \ass@sfont{m@thsy}{\fontn@me}%
    \get@font{mthsy}{\math@syfnt}{\sss@ze}\set@skchar{'60}%
    \ass@ssfont{m@thsy}{\fontn@me}%
    \math@atom{#1}{%
    \mathchoice%
      {\hbox{$\m@th\displaystyle#1$}}%
      {\hbox{$\m@th\textstyle#1$}}%
      {\hbox{$\m@th\scriptstyle#1$}}%
      {\hbox{$\m@th\scriptscriptstyle#1$}}}%
  \endgroup
  }%
}



\def\diameter{{\ifmmode\mathchoice
{\ooalign{\hfil\hbox{$\displaystyle/$}\hfil\crcr
{\hbox{$\displaystyle\mathchar"20D$}}}}
{\ooalign{\hfil\hbox{$\textstyle/$}\hfil\crcr
{\hbox{$\textstyle\mathchar"20D$}}}}
{\ooalign{\hfil\hbox{$\scriptstyle/$}\hfil\crcr
{\hbox{$\scriptstyle\mathchar"20D$}}}}
{\ooalign{\hfil\hbox{$\scriptscriptstyle/$}\hfil\crcr
{\hbox{$\scriptscriptstyle\mathchar"20D$}}}}
\else{\ooalign{\hfil/\hfil\crcr\mathhexbox20D}}%
\fi}}

\def\sq{\ifmmode\squareforqed\else{\unskip\nobreak\hfil
\penalty50\hskip1em\null\nobreak\hfil\squareforqed
\parfillskip=0pt\finalhyphendemerits=0\endgraf}\fi}
\def\squareforqed{\hbox{\rlap{$\sqcap$}$\sqcup$}}


\def\bbbc{{\mathchoice {\setbox0=\hbox{$\displaystyle\rm C$}\hbox{\hbox
to0pt{\kern0.4\wd0\vrule height0.9\ht0\hss}\box0}}
{\setbox0=\hbox{$\textstyle\rm C$}\hbox{\hbox
to0pt{\kern0.4\wd0\vrule height0.9\ht0\hss}\box0}}
{\setbox0=\hbox{$\scriptstyle\rm C$}\hbox{\hbox
to0pt{\kern0.4\wd0\vrule height0.9\ht0\hss}\box0}}
{\setbox0=\hbox{$\scriptscriptstyle\rm C$}\hbox{\hbox
to0pt{\kern0.4\wd0\vrule height0.9\ht0\hss}\box0}}}}
\def\bbbq{{\mathchoice {\setbox0=\hbox{$\displaystyle\rm
Q$}\hbox{\raise
0.15\ht0\hbox to0pt{\kern0.4\wd0\vrule height0.8\ht0\hss}\box0}}
{\setbox0=\hbox{$\textstyle\rm Q$}\hbox{\raise
0.15\ht0\hbox to0pt{\kern0.4\wd0\vrule height0.8\ht0\hss}\box0}}
{\setbox0=\hbox{$\scriptstyle\rm Q$}\hbox{\raise
0.15\ht0\hbox to0pt{\kern0.4\wd0\vrule height0.7\ht0\hss}\box0}}
{\setbox0=\hbox{$\scriptscriptstyle\rm Q$}\hbox{\raise
0.15\ht0\hbox to0pt{\kern0.4\wd0\vrule height0.7\ht0\hss}\box0}}}}
\def\bbbt{{\mathchoice {\setbox0=\hbox{$\displaystyle\rm
T$}\hbox{\hbox to0pt{\kern0.3\wd0\vrule height0.9\ht0\hss}\box0}}
{\setbox0=\hbox{$\textstyle\rm T$}\hbox{\hbox
to0pt{\kern0.3\wd0\vrule height0.9\ht0\hss}\box0}}
{\setbox0=\hbox{$\scriptstyle\rm T$}\hbox{\hbox
to0pt{\kern0.3\wd0\vrule height0.9\ht0\hss}\box0}}
{\setbox0=\hbox{$\scriptscriptstyle\rm T$}\hbox{\hbox
to0pt{\kern0.3\wd0\vrule height0.9\ht0\hss}\box0}}}}
\def\bbbs{{\mathchoice
{\setbox0=\hbox{$\displaystyle     \rm S$}\hbox{\raise0.5\ht0\hbox
to0pt{\kern0.35\wd0\vrule height0.45\ht0\hss}\hbox
to0pt{\kern0.55\wd0\vrule height0.5\ht0\hss}\box0}}
{\setbox0=\hbox{$\textstyle        \rm S$}\hbox{\raise0.5\ht0\hbox
to0pt{\kern0.35\wd0\vrule height0.45\ht0\hss}\hbox
to0pt{\kern0.55\wd0\vrule height0.5\ht0\hss}\box0}}
{\setbox0=\hbox{$\scriptstyle      \rm S$}\hbox{\raise0.5\ht0\hbox
to0pt{\kern0.35\wd0\vrule height0.45\ht0\hss}\raise0.05\ht0\hbox
to0pt{\kern0.5\wd0\vrule height0.45\ht0\hss}\box0}}
{\setbox0=\hbox{$\scriptscriptstyle\rm S$}\hbox{\raise0.5\ht0\hbox
to0pt{\kern0.4\wd0\vrule height0.45\ht0\hss}\raise0.05\ht0\hbox
to0pt{\kern0.55\wd0\vrule height0.45\ht0\hss}\box0}}}}
\def\bbbz{{\mathchoice {\hbox{$\sf\textstyle Z\kern-0.4em Z$}}
{\hbox{$\sf\textstyle Z\kern-0.4em Z$}}
{\hbox{$\sf\scriptstyle Z\kern-0.3em Z$}}
{\hbox{$\sf\scriptscriptstyle Z\kern-0.2em Z$}}}}


\def\Nulle{0} 
\def\Afe{1}   
\def\Hae{2}   
\def\Hbe{3}   
\def\Hce{4}   
\def\Hde{5}   


\newcount\LastMac       \LastMac=\Nulle

\newskip\half      \half=5.5pt plus 1.5pt minus 2.25pt
\newskip\one       \one=11pt plus 3pt minus 5.5pt
\newskip\onehalf   \onehalf=16.5pt plus 5.5pt minus 8.25pt
\newskip\two       \two=22pt plus 5.5pt minus 11pt

\def\Half{\addvspace{\half}}
\def\One{\addvspace{\one}}
\def\OneHalf{\addvspace{\onehalf}}
\def\Two{\addvspace{\two}}

\def\Raggedright{
  \rightskip=\z@ plus \hsize\relax
}

\def\Fullout{
  \rightskip=\z@\relax
}

\def\Hang#1#2{
  \hangindent=#1%
  \hangafter=#2\relax
}


\newif\ifsp@page
\def\pagestyle#1{\csname ps@#1\endcsname}
\def\thispagestyle#1{\global\sp@pagetrue\gdef\sp@type{#1}}

\def\ps@titlepage{%
  \def\@oddhead{\eightpoint\noindent \the\CatchLine
    \ifprod@font\else\qquad Printed\ \today\qquad
      (MN plain \TeX\ macros\ v\@version)\fi \hfil}%
  \let\@evenhead=\@oddhead
  \def\@oddfoot{\eightpoint\copyright\ \@pubyear\ RAS\hfil}%
  \def\@evenfoot{\hfil\eightpoint\noindent\copyright\ \@pubyear\ RAS}%
}

\def\ps@headings{%
  \def\@oddhead{\elevenpoint\it\noindent
    \hfill\the\RightHeader\hskip1.5em\rm\folio}%
  \def\@evenhead{\elevenpoint\noindent
    \folio\hskip1.5em\it\the\LeftHeader\hfill}%
  \def\@oddfoot{\eightpoint\noindent\copyright\ \@pubyear\ RAS,
    MNRAS {\bf \@volume}, \@pagerange\hfil}%
  \def\@evenfoot{\hfil\eightpoint\copyright\ \@pubyear\ RAS,
    MNRAS {\bf \@volume}, \@pagerange}%
}

\def\ps@plate{%
  \def\@oddhead{\eightpoint\noindent\plt@cap\hfil}%
  \def\@evenhead{\eightpoint\noindent\plt@cap\hfil}%
  \def\@oddfoot{\eightpoint\noindent\copyright\ \@pubyear\ RAS,
    MNRAS {\bf \@volume}, \@pagerange\hfil}%
  \def\@evenfoot{\hfil\eightpoint\copyright\ \@pubyear\ RAS,
    MNRAS {\bf \@volume}, \@pagerange}%
}



\def\title#1{
  \bgroup
    \vbox to 8pt{\vss}%
    \seventeenpoint
    \Raggedright
    \noindent \strut{\bf #1}\par
  \egroup
}

\def\author#1{
  \bgroup
    \ifnum\LastMac=\Afe \OneHalf\else \vskip 21pt\fi
    \fourteenpoint
    \Raggedright
    \noindent \strut #1\par
    \vskip 3pt%
  \egroup
}

\def\affiliation#1{
  \bgroup
    \vskip -4pt%
    \eightpoint
    \Raggedright
    \noindent \strut {\it #1}\par
  \egroup
  \LastMac=\Afe\relax
}

\def\acceptedline#1{
  \bgroup
    \Two
    \eightpoint
    \Raggedright
    \noindent \strut #1\par
  \egroup
}

\long\def\abstract#1{%
  \bgroup
    \vskip 20pt%
    \leftskip 11pc\rightskip\z@
    \noindent{\ninebf ABSTRACT}\par
    \tenpoint
    \Fullout
    \noindent #1\par
  \egroup
}

\long\def\keywords#1{
  \bgroup
    \Half
    \leftskip 11pc\rightskip\z@
    \tenpoint
    \Fullout
    \noindent\hbox{\bf Key words:}\ #1\par
  \egroup
}


\def\maketitle{%
  \EndOpening
  \ifsinglecol \else \MakePage\fi
}


\def\pageoffset#1#2{\hoffset=#1\relax\voffset=#2\relax}


\def\@nameuse#1{\csname #1\endcsname}
\def\arabic#1{\@arabic{\@nameuse{#1}}}
\def\alph#1{\@alph{\@nameuse{#1}}}
\def\Alph#1{\@Alph{\@nameuse{#1}}}
\def\@arabic#1{\number #1}
\def\@Alph#1{\ifcase#1\or A\or B\or C\or D\else\@Ialph{#1}\fi}
\def\@Ialph#1{\ifcase#1\or \or \or \or \or E\or F\or G\or H\or I\or J\or
   K\or L\or M\or N\or O\or P\or Q\or R\or S\or T\or U\or V\or W\or X\or
   Y\or Z\else\errmessage{Counter out of range}\fi}
\def\@alph#1{\ifcase#1\or a\or b\or c\or d\else\@ialph{#1}\fi}
\def\@ialph#1{\ifcase#1\or \or \or \or \or e\or f\or g\or h\or i\or j\or
   k\or l\or m\or n\or o\or p\or q\or r\or s\or t\or u\or v\or w\or x\or y\or
   z\else\errmessage{Counter out of range}\fi}


\newcount\Eqnno
\newcount\SubEqnno

\def\theeq{\arabic{Eqnno}}
\def\thesubeq{\alph{SubEqnno}}

\def\stepeq{\relax
  \global\SubEqnno \z@
  \global\advance\Eqnno \@ne\relax
  {\rm (\theeq)}%
}

\def\startsubeq{\relax
  \global\SubEqnno \z@
  \global\advance\Eqnno \@ne\relax
  \stepsubeq
}

\def\stepsubeq{\relax
  \global\advance\SubEqnno \@ne\relax
  {\rm (\theeq\thesubeq)}%
}


\newcount\Sec        
\newcount\SecSec
\newcount\SecSecSec

\def\thesection{\arabic{Sec}}
\def\thesubsection{\thesection.\arabic{SecSec}}
\def\thesubsubsection{\thesubsection.\arabic{SecSecSec}}

\Sec=\z@

\def\:{\let\@sptoken= } \:  
\def\:{\@xifnch} \expandafter\def\: {\futurelet\@tempc\@ifnch}

\def\@ifnextchar#1#2#3{%
  \let\@tempMACe #1%
  \def\@tempMACa{#2}%
  \def\@tempMACb{#3}%
  \futurelet \@tempMACc\@ifnch%
}

\def\@ifnch{%
\ifx \@tempMACc \@sptoken%
  \let\@tempMACd\@xifnch%
\else%
  \ifx \@tempMACc \@tempMACe%
    \let\@tempMACd\@tempMACa%
  \else%
    \let\@tempMACd\@tempMACb%
  \fi%
\fi%
\@tempMACd%
}

\def\@ifstar#1#2{\@ifnextchar *{\def\@tempMACa*{#1}\@tempMACa}{#2}}

\newskip\@tempskipb

\def\addvspace#1{%
  \ifvmode\else \endgraf\fi%
  \ifdim\lastskip=\z@%
    \vskip #1\relax%
  \else%
    \@tempskipb#1\relax\@xaddvskip%
  \fi%
}

\def\@xaddvskip{%
  \ifdim\lastskip<\@tempskipb%
    \vskip-\lastskip%
    \vskip\@tempskipb\relax%
  \else%
    \ifdim\@tempskipb<\z@%
      \ifdim\lastskip<\z@ \else%
        \advance\@tempskipb\lastskip%
        \vskip-\lastskip\vskip\@tempskipb%
      \fi%
    \fi%
  \fi%
}

\newskip\@tmpSKIP

\def\addpen#1{%
  \ifvmode
    \if@nobreak
    \else
      \ifdim\lastskip=\z@
        \penalty#1\relax
      \else
        \@tmpSKIP=\lastskip
        \vskip -\lastskip
        \penalty#1\vskip\@tmpSKIP
      \fi
    \fi
  \fi
}

\newcount\@clubpen   \@clubpen=\clubpenalty
\newif\if@nobreak    \@nobreakfalse

\def\@noafterindent{%
  \global\@nobreaktrue
  \everypar{\if@nobreak
              \global\@nobreakfalse
              \clubpenalty \@M
              {\setbox\z@\lastbox}%
              \LastMac=\Nulle\relax%
            \else
              \clubpenalty \@clubpen
              \everypar{}%
            \fi}%
}

\newcount\gds@cbrk   \gds@cbrk=-300

\def\@nohdbrk{\interlinepenalty \@M\relax}

\let\@par=\par
\def\@restorepar{\def\par{\@par}}

\newif\if@endpe   \@endpefalse
 
\def\@doendpe{\@endpetrue \@nobreakfalse \LastMac=\Nulle\relax%
     \def\par{\@restorepar\everypar{}\par\@endpefalse}%
              \everypar{\setbox\z@\lastbox\everypar{}\@endpefalse}%
}

\def\section{\@ifstar{\@ssection}{\@section}}

\def\@section#1{
  \if@nobreak
    \everypar{}%
    \ifnum\LastMac=\Hae \addvspace{\half}\fi
  \else
    \addpen{\gds@cbrk}%
    \addvspace{\two}%
  \fi
  \bgroup
    \ninepoint\bf
    \Raggedright
    \global\advance\Sec \@ne
    \ifappendix
      \global\Eqnno=\z@ \global\SubEqnno=\z@\relax
      \def\ch@ck{#1}%
      \ifx\ch@ck\empty \def\c@lon{}\else\def\c@lon{:}\fi
      \noindent\@nohdbrk APPENDIX\ \thesection\c@lon\hskip 0.5em%
        \uppercase{#1}\par
    \else
      \noindent\@nohdbrk\thesection\hskip 1pc \uppercase{#1}\par
    \fi
    \global\SecSec=\z@
  \egroup
  \nobreak
  \vskip\half
  \nobreak
  \@noafterindent
  \LastMac=\Hae\relax
}

\def\@ssection#1{
  \if@nobreak
    \everypar{}%
    \ifnum\LastMac=\Hae \addvspace{\half}\fi
  \else
    \addpen{\gds@cbrk}%
    \addvspace{\two}%
  \fi
  \bgroup
    \ninepoint\bf
    \Raggedright
    \noindent\@nohdbrk\uppercase{#1}\par
  \egroup
  \nobreak
  \vskip\half
  \nobreak
  \@noafterindent
  \LastMac=\Hae\relax
}

\def\subsection{\@ifstar{\@ssubsection}{\@subsection}}

\def\@subsection#1{
  \if@nobreak
    \everypar{}%
    \ifnum\LastMac=\Hae \addvspace{1pt plus 1pt minus .5pt}\fi
  \else
    \addpen{\gds@cbrk}%
    \addvspace{\onehalf}%
  \fi
  \bgroup
    \ninepoint\bf
    \Raggedright
    \global\advance\SecSec \@ne
    \noindent\@nohdbrk\thesubsection \hskip 1pc\relax #1\par
    \global\SecSecSec=\z@
  \egroup
  \nobreak
  \vskip\half
  \nobreak
  \@noafterindent
  \LastMac=\Hbe\relax
}

\def\@ssubsection#1{
  \if@nobreak
    \everypar{}%
    \ifnum\LastMac=\Hae \addvspace{1pt plus 1pt minus .5pt}\fi
  \else
    \addpen{\gds@cbrk}%
    \addvspace{\onehalf}%
  \fi
  \bgroup
    \ninepoint\bf
    \Raggedright
    \noindent\@nohdbrk #1\par
  \egroup
  \nobreak
  \vskip\half
  \nobreak
  \@noafterindent
  \LastMac=\Hbe\relax
}

\def\subsubsection{\@ifstar{\@ssubsubsection}{\@subsubsection}}

\def\@subsubsection#1{
  \if@nobreak
    \everypar{}%
    \ifnum\LastMac=\Hbe \addvspace{1pt plus 1pt minus .5pt}\fi
  \else
    \addpen{\gds@cbrk}%
    \addvspace{\onehalf}%
  \fi
  \bgroup
    \ninepoint\it
    \Raggedright
    \global\advance\SecSecSec \@ne
    \noindent\@nohdbrk\thesubsubsection \hskip 1pc\relax #1\par
  \egroup
  \nobreak
  \vskip\half
  \nobreak
  \@noafterindent
  \LastMac=\Hce\relax
}

\def\@ssubsubsection#1{
  \if@nobreak
    \everypar{}%
    \ifnum\LastMac=\Hbe \addvspace{1pt plus 1pt minus .5pt}\fi
  \else
    \addpen{\gds@cbrk}%
    \addvspace{\onehalf}%
  \fi
  \bgroup
    \ninepoint\it
    \Raggedright
    \noindent\@nohdbrk #1\par
  \egroup
  \nobreak
  \vskip\half
  \nobreak
  \@noafterindent
  \LastMac=\Hce\relax
}

\def\paragraph#1{
  \if@nobreak
    \everypar{}%
  \else
    \addpen{\gds@cbrk}%
    \addvspace{\one}%
  \fi%
  \bgroup%
    \ninepoint\it
    \noindent #1\ \nobreak%
  \egroup
  \LastMac=\Hde\relax
  \ignorespaces
}


\newif\ifappendix

\def\appendix{%
  \global\appendixtrue
  \def\thesection{\Alph{Sec}}%
  \def\thesubsection{\thesection\arabic{SecSec}}%
  \def\theeq{\thesection\arabic{Eqnno}}%
  \Sec=\z@ \SecSec=\z@ \SecSecSec=\z@ \Eqnno=\z@ \SubEqnno=\z@\relax
}




\def\beginlist{%
  \par\if@nobreak \else\addvspace{\half}\fi%
  \bgroup%
    \ninepoint
    \let\item=\list@item%
}

\def\list@item{%
  \par\noindent\hskip 1em\relax%
  \ignorespaces%
}

\def\endlist{\par\egroup\addvspace{\half}\@doendpe}


\def\beginrefs{%
  \par
  \bgroup
    \eightpoint
    \Fullout
    \let\bibitem=\bib@item
}

\def\bib@item{%
  \par\parindent=1.5em\Hang{1.5em}{1}%
  \everypar={\Hang{1.5em}{1}\ignorespaces}%
  \noindent\ignorespaces
}

\def\endrefs{\par\egroup\@doendpe}


\newtoks\CatchLine

\def\@journal{Mon.\ Not.\ R.\ Astron.\ Soc.\ }  
\def\@pubyear{1997}        
\def\@pagerange{000--000}  
\def\@volume{000}          
\def\@microfiche{}         %

\def\pubyear#1{\gdef\@pubyear{#1}\@makecatchline}
\def\pagerange#1{\gdef\@pagerange{#1}\@makecatchline}
\def\volume#1{\gdef\@volume{#1}\@makecatchline}
\def\microfiche#1{\gdef\@microfiche{and Microfiche\ #1}\@makecatchline}

\def\@makecatchline{%
  \global\CatchLine{%
    {\rm \@journal {\bf \@volume},\ \@pagerange\ (\@pubyear)\ \@microfiche}}%
}

\@makecatchline 

\newtoks\LeftHeader
\def\shortauthor#1{
  \global\LeftHeader{#1}%
}

\newtoks\RightHeader
\def\shorttitle#1{
  \global\RightHeader{#1}%
}

\def\PageHead{
  \begingroup
    \ifsp@page
      \csname ps@\sp@type\endcsname
    \fi
    \ifodd\pageno
      \let\the@head=\@oddhead
    \else
      \let\the@head=\@evenhead
    \fi
    \vbox to \z@{\vskip-22.5\p@%
      \hbox to \PageWidth{\vbox to8.5\p@{}%
        \the@head
      }%
    \vss}%
  \endgroup
  \nointerlineskip
}

\gdef\PageFoot{%
  \nointerlineskip%
  \begingroup
  \ifsp@page
    \csname ps@\sp@type\endcsname
    \global\sp@pagefalse
  \fi
  \vbox to 22pt{\vfil%
    \hbox to \PageWidth{%
      \eightpoint\strut\noindent
      \ifodd\pageno
        \@oddfoot
      \else
        \@evenfoot
      \fi
    }%
  }%
  \endgroup
}

\def\today{%
  \number\day\space
  \ifcase\month\or January\or February\or March\or April\or May\or June\or
    July\or August\or September\or October\or November\or December\fi
  \space\number\year%
}

\def\authorcomment#1{%
  \gdef\PageFoot{%
    \nointerlineskip%
    \vbox to 20pt{\vfil%
      \hbox to \PageWidth{\elevenpoint\noindent \hfil #1 \hfil}}%
  }%
}


\newif\ifplate@page
\newbox\plt@box

\def\beginplatepage{%
  \let\plate=\plate@head
  \let\caption=\fig@caption
  \global\setbox\plt@box=\vbox\bgroup
  \TEMPDIMEN=\PageWidth 
  \hsize=\PageWidth\relax
}

\def\endplatepage{\par\egroup\global\plate@pagetrue}
\def\plate@head#1{\gdef\plt@cap{#1}}


\def\letters{%
  \gdef\folio{\ifnum\pageno<\z@ L\romannumeral-\pageno
    \else L\number\pageno \fi}%
}


\newdimen\mathindent

\global\mathindent=\z@
\global\everydisplay{\global\@dspwd=\displaywidth\displaysetup}


\def\@displaylines#1{
  {}$\displ@y\hbox{\vbox{\halign{$\@lign\hfil\displaystyle##\hfil$\crcr
  #1\crcr}}}${}%
}

\def\@eqalign#1{\null\vcenter{\openup\jot\m@th
  \ialign{\strut\hfil$\displaystyle{##}$&$\displaystyle{{}##}$\hfil
      \crcr#1\crcr}}%
}

\def\@eqalignno#1{
  \global\advance\@dspwd by -\mathindent%
  {}$\displ@y\hbox{\vbox{\halign to\@dspwd%
  {\hfil$\@lign\displaystyle{##}$\tabskip\z@skip
  &$\@lign\displaystyle{{}##}$\hfil\tabskip\centering
  &\llap{$\@lign##$}\tabskip\z@skip\crcr
  #1\crcr}}}${}%
}


\global\let\displaylines=\@displaylines
\global\let\eqalign=\@eqalign
\global\let\eqalignno=\@eqalignno
\global\let\leqalignno=\@eqalignno

\newdimen\@dspwd   \@dspwd=\z@
\newif\if@eqno
\newif\if@leqno
\newtoks\@eqn
\newtoks\@eq

\def\displaysetup#1$${\displaytest#1\eqno\eqno\displaytest}

\def\displaytest#1\eqno#2\eqno#3\displaytest{%
 \if!#3!\ldisplaytest#1\leqno\leqno\ldisplaytest
 \else\@eqnotrue\@leqnofalse\@eqn={#2}\@eq={#1}\fi
 \generaldisplay$$}

\def\ldisplaytest#1\leqno#2\leqno#3\ldisplaytest{%
\@eq={#1}%
 \if!#3!\@eqnofalse\else\@eqnotrue\@leqnotrue
  \@eqn={#2}\fi}

\def\generaldisplay{%
  \if@eqno
    \if@leqno
      \hbox to \displaywidth{\noindent
        \rlap{$\displaystyle\the\@eqn$}%
        \hskip\mathindent$\displaystyle\the\@eq$\hfil}%
    \else
      \hbox to \displaywidth{\noindent
        \hskip\mathindent
        $\displaystyle\the\@eq$\hfil$\displaystyle\the\@eqn$}%
    \fi
  \else
    \hbox to \displaywidth{\noindent
      \hskip\mathindent$\displaystyle\the\@eq$\hfil}%
  \fi
}


\def\@notice{%
  \par\Two%
  \noindent{\b@ls{11pt}\ninerm This paper has been produced using the
    Royal Astronomical Society/Blackwell Science \TeX\ macros.\par}%
}

\outer\def\bye{\@notice\par\vfill\supereject\end}


\def\start@mess{%
  Monthly notices of the RAS journal style (\@typeface)\space
    v\@version,\space \@verdate.%
}

\everyjob{\Warn{\start@mess}}



\newif\if@debug \@debugfalse  

\def\Print#1{\if@debug\immediate\write16{#1}\else \fi}
\def\Warn#1{\immediate\write16{#1}}
\def\wlog#1{}

\newcount\Iteration 

\def\Single{0} \def\Double{1}                 
\def\Figure{0} \def\Table{1}                  

\def\InStack{0}  
\def\InZoneA{1}
\def\InZoneB{2}
\def\InZoneC{3}

\newcount\TEMPCOUNT 
\newdimen\TEMPDIMEN 
\newbox\TEMPBOX     
\newbox\VOIDBOX     

\newcount\LengthOfStack 
\newcount\MaxItems      
\newcount\StackPointer
\newcount\Point         
\newcount\NextFigure    
\newcount\NextTable     
\newcount\NextItem      

\newcount\StatusStack   
\newcount\NumStack      
\newcount\TypeStack     
\newcount\SpanStack     
\newcount\BoxStack      

\newcount\ItemSTATUS    
\newcount\ItemNUMBER    
\newcount\ItemTYPE      
\newcount\ItemSPAN      
\newbox\ItemBOX         
\newdimen\ItemSIZE      

\newdimen\PageHeight    
\newdimen\TextLeading   
\newdimen\Feathering    
\newcount\LinesPerPage  
\newdimen\ColumnWidth   
\newdimen\ColumnGap     
\newdimen\PageWidth     
\newdimen\BodgeHeight   
\newcount\Leading       

\newdimen\ZoneBSize  
\newdimen\TextSize   
\newbox\ZoneABOX     
\newbox\ZoneBBOX     
\newbox\ZoneCBOX     

\newif\ifFirstSingleItem
\newif\ifFirstZoneA
\newif\ifMakePageInComplete
\newif\ifMoreFigures \MoreFiguresfalse 
\newif\ifMoreTables  \MoreTablesfalse  

\newif\ifFigInZoneB 
\newif\ifFigInZoneC 
\newif\ifTabInZoneB 
\newif\ifTabInZoneC

\newif\ifZoneAFullPage

\newbox\MidBOX    
\newbox\LeftBOX
\newbox\RightBOX
\newbox\PageBOX   

\newif\ifLeftCOL  
\LeftCOLtrue

\newdimen\ZoneBAdjust

\newcount\ItemFits
\def\Yes{1}
\def\No{2}


\MaxItems=15
\NextFigure=\z@        
\NextTable=\@ne

\BodgeHeight=6pt
\TextLeading=11pt    
\Leading=11
\Feathering=\z@      
\LinesPerPage=61     
\topskip=\TextLeading
\ColumnWidth=20pc    
\ColumnGap=2pc       

\newskip\ItemSepamount  
\ItemSepamount=\TextLeading plus \TextLeading minus 4pt

\parskip=\z@ plus .1pt
\parindent=18pt
\widowpenalty=\z@
\clubpenalty=10000
\tolerance=1500
\hbadness=1500
\abovedisplayskip=6pt plus 2pt minus 1pt
\belowdisplayskip=6pt plus 2pt minus 1pt
\abovedisplayshortskip=6pt plus 2pt minus 1pt
\belowdisplayshortskip=6pt plus 2pt minus 1pt

\frenchspacing

\ninepoint 

\PageHeight=682pt
\PageWidth=2\ColumnWidth
\advance\PageWidth by \ColumnGap

\pagestyle{headings}




\newcount\DUMMY \StatusStack=\allocationnumber
\newcount\DUMMY \newcount\DUMMY \newcount\DUMMY 
\newcount\DUMMY \newcount\DUMMY \newcount\DUMMY 
\newcount\DUMMY \newcount\DUMMY \newcount\DUMMY
\newcount\DUMMY \newcount\DUMMY \newcount\DUMMY 
\newcount\DUMMY \newcount\DUMMY \newcount\DUMMY

\newcount\DUMMY \NumStack=\allocationnumber
\newcount\DUMMY \newcount\DUMMY \newcount\DUMMY 
\newcount\DUMMY \newcount\DUMMY \newcount\DUMMY 
\newcount\DUMMY \newcount\DUMMY \newcount\DUMMY 
\newcount\DUMMY \newcount\DUMMY \newcount\DUMMY 
\newcount\DUMMY \newcount\DUMMY \newcount\DUMMY

\newcount\DUMMY \TypeStack=\allocationnumber
\newcount\DUMMY \newcount\DUMMY \newcount\DUMMY 
\newcount\DUMMY \newcount\DUMMY \newcount\DUMMY 
\newcount\DUMMY \newcount\DUMMY \newcount\DUMMY 
\newcount\DUMMY \newcount\DUMMY \newcount\DUMMY 
\newcount\DUMMY \newcount\DUMMY \newcount\DUMMY

\newcount\DUMMY \SpanStack=\allocationnumber
\newcount\DUMMY \newcount\DUMMY \newcount\DUMMY 
\newcount\DUMMY \newcount\DUMMY \newcount\DUMMY 
\newcount\DUMMY \newcount\DUMMY \newcount\DUMMY 
\newcount\DUMMY \newcount\DUMMY \newcount\DUMMY 
\newcount\DUMMY \newcount\DUMMY \newcount\DUMMY

\newbox\DUMMY   \BoxStack=\allocationnumber
\newbox\DUMMY   \newbox\DUMMY \newbox\DUMMY 
\newbox\DUMMY   \newbox\DUMMY \newbox\DUMMY 
\newbox\DUMMY   \newbox\DUMMY \newbox\DUMMY 
\newbox\DUMMY   \newbox\DUMMY \newbox\DUMMY 
\newbox\DUMMY   \newbox\DUMMY \newbox\DUMMY

\def\wlog{\immediate\write\m@ne}


\def\GetItemAll#1{%
 \GetItemSTATUS{#1}
 \GetItemNUMBER{#1}
 \GetItemTYPE{#1}
 \GetItemSPAN{#1}
 \GetItemBOX{#1}
}

\def\GetItemSTATUS#1{%
 \Point=\StatusStack
 \advance\Point by #1
 \global\ItemSTATUS=\count\Point
}

\def\GetItemNUMBER#1{%
 \Point=\NumStack
 \advance\Point by #1
 \global\ItemNUMBER=\count\Point
}

\def\GetItemTYPE#1{%
 \Point=\TypeStack
 \advance\Point by #1
 \global\ItemTYPE=\count\Point
}

\def\GetItemSPAN#1{%
 \Point\SpanStack
 \advance\Point by #1
 \global\ItemSPAN=\count\Point
}

\def\GetItemBOX#1{%
 \Point=\BoxStack
 \advance\Point by #1
 \global\setbox\ItemBOX=\vbox{\copy\Point}
 \global\ItemSIZE=\ht\ItemBOX
 \global\advance\ItemSIZE by \dp\ItemBOX
 \TEMPCOUNT=\ItemSIZE
 \divide\TEMPCOUNT by \Leading
 \divide\TEMPCOUNT by 65536
 \advance\TEMPCOUNT \@ne
 \ItemSIZE=\TEMPCOUNT pt
 \global\multiply\ItemSIZE by \Leading
}


\def\JoinStack{%
 \ifnum\LengthOfStack=\MaxItems 
  \Warn{WARNING: Stack is full...some items will be lost!}
 \else
  \Point=\StatusStack
  \advance\Point by \LengthOfStack
  \global\count\Point=\ItemSTATUS
  \Point=\NumStack
  \advance\Point by \LengthOfStack
  \global\count\Point=\ItemNUMBER
  \Point=\TypeStack
  \advance\Point by \LengthOfStack
  \global\count\Point=\ItemTYPE
  \Point\SpanStack
  \advance\Point by \LengthOfStack
  \global\count\Point=\ItemSPAN
  \Point=\BoxStack
  \advance\Point by \LengthOfStack
  \global\setbox\Point=\vbox{\copy\ItemBOX}
  \global\advance\LengthOfStack \@ne
  \ifnum\ItemTYPE=\Figure 
   \global\MoreFigurestrue
  \else
   \global\MoreTablestrue
  \fi
 \fi
}


\def\LeaveStack#1{%
 {\Iteration=#1
 \loop
 \ifnum\Iteration<\LengthOfStack
  \advance\Iteration \@ne
  \GetItemSTATUS{\Iteration}
   \advance\Point by \m@ne
   \global\count\Point=\ItemSTATUS
  \GetItemNUMBER{\Iteration}
   \advance\Point by \m@ne
   \global\count\Point=\ItemNUMBER
  \GetItemTYPE{\Iteration}
   \advance\Point by \m@ne
   \global\count\Point=\ItemTYPE
  \GetItemSPAN{\Iteration}
   \advance\Point by \m@ne
   \global\count\Point=\ItemSPAN
  \GetItemBOX{\Iteration}
   \advance\Point by \m@ne
   \global\setbox\Point=\vbox{\copy\ItemBOX}
 \repeat}
 \global\advance\LengthOfStack by \m@ne
}


\newif\ifStackNotClean

\def\CleanStack{%
 \StackNotCleantrue
 {\Iteration=\z@
  \loop
   \ifStackNotClean
    \GetItemSTATUS{\Iteration}
    \ifnum\ItemSTATUS=\InStack
     \advance\Iteration \@ne
     \else
      \LeaveStack{\Iteration}
    \fi
   \ifnum\LengthOfStack<\Iteration
    \StackNotCleanfalse
   \fi
 \repeat}
}


\def\FindItem#1#2{%
 \global\StackPointer=\m@ne 
 {\Iteration=\z@
  \loop
  \ifnum\Iteration<\LengthOfStack
   \GetItemSTATUS{\Iteration}
   \ifnum\ItemSTATUS=\InStack
    \GetItemTYPE{\Iteration}
    \ifnum\ItemTYPE=#1
     \GetItemNUMBER{\Iteration}
     \ifnum\ItemNUMBER=#2
      \global\StackPointer=\Iteration
      \Iteration=\LengthOfStack 
     \fi
    \fi
   \fi
  \advance\Iteration \@ne
 \repeat}
}


\def\FindNext{%
 \global\StackPointer=\m@ne 
 {\Iteration=\z@
  \loop
  \ifnum\Iteration<\LengthOfStack
   \GetItemSTATUS{\Iteration}
   \ifnum\ItemSTATUS=\InStack
    \GetItemTYPE{\Iteration}
   \ifnum\ItemTYPE=\Figure
    \ifMoreFigures
      \global\NextItem=\Figure
      \global\StackPointer=\Iteration
      \Iteration=\LengthOfStack 
    \fi
   \fi
   \ifnum\ItemTYPE=\Table
    \ifMoreTables
      \global\NextItem=\Table
      \global\StackPointer=\Iteration
      \Iteration=\LengthOfStack 
    \fi
   \fi
  \fi
  \advance\Iteration \@ne
 \repeat}
}


\def\ChangeStatus#1#2{%
 \Point=\StatusStack
 \advance\Point by #1
 \global\count\Point=#2
}



\def\Zone{\InZoneA}

\ZoneBAdjust=\z@

\def\MakePage{
 \global\ZoneBSize=\PageHeight
 \global\TextSize=\ZoneBSize
 \global\ZoneAFullPagefalse
 \global\topskip=\TextLeading
 \MakePageInCompletetrue
 \MoreFigurestrue
 \MoreTablestrue
 \FigInZoneBfalse
 \FigInZoneCfalse
 \TabInZoneBfalse
 \TabInZoneCfalse
 \global\FirstSingleItemtrue
 \global\FirstZoneAtrue
 \global\setbox\ZoneABOX=\box\VOIDBOX
 \global\setbox\ZoneBBOX=\box\VOIDBOX
 \global\setbox\ZoneCBOX=\box\VOIDBOX
 \loop
  \ifMakePageInComplete
 \FindNext
 \ifnum\StackPointer=\m@ne
  \NextItem=\m@ne
  \MoreFiguresfalse
  \MoreTablesfalse
 \fi
 \ifnum\NextItem=\Figure
   \FindItem{\Figure}{\NextFigure}
   \ifnum\StackPointer=\m@ne \global\MoreFiguresfalse
   \else
    \GetItemSPAN{\StackPointer}
    \ifnum\ItemSPAN=\Single \def\Zone{\InZoneB}\relax
     \ifFigInZoneC \global\MoreFiguresfalse\fi
    \else
     \def\Zone{\InZoneA}
     \ifFigInZoneB \def\Zone{\InZoneC}\fi
    \fi
   \fi
   \ifMoreFigures\Print{}\FigureItems\fi
 \fi
\ifnum\NextItem=\Table
   \FindItem{\Table}{\NextTable}
   \ifnum\StackPointer=\m@ne \global\MoreTablesfalse
   \else
    \GetItemSPAN{\StackPointer}
    \ifnum\ItemSPAN=\Single\relax
     \ifTabInZoneC \global\MoreTablesfalse\fi
    \else
     \def\Zone{\InZoneA}
     \ifTabInZoneB \def\Zone{\InZoneC}\fi
    \fi
   \fi
   \ifMoreTables\Print{}\TableItems\fi
 \fi
   \MakePageInCompletefalse 
   \ifMoreFigures\MakePageInCompletetrue\fi
   \ifMoreTables\MakePageInCompletetrue\fi
 \repeat
 \ifZoneAFullPage
  \global\TextSize=\z@
  \global\ZoneBSize=\z@
  \global\vsize=\z@\relax
  \global\topskip=\z@\relax
  \vbox to \z@{\vss}
  \eject
 \else
 \global\advance\ZoneBSize by -\ZoneBAdjust
 \global\vsize=\ZoneBSize
 \global\hsize=\ColumnWidth
 \global\ZoneBAdjust=\z@
 \ifdim\TextSize<23pt
 \Warn{}
 \Warn{* Making column fall short: TextSize=\the\TextSize *}
 \vskip-\lastskip\eject\fi
 \fi
}

\def\MakeRightCol{
 \global\TextSize=\ZoneBSize
 \MakePageInCompletetrue
 \MoreFigurestrue
 \MoreTablestrue
 \global\FirstSingleItemtrue
 \global\setbox\ZoneBBOX=\box\VOIDBOX
 \def\Zone{\InZoneB}
 \loop
  \ifMakePageInComplete
 \FindNext
 \ifnum\StackPointer=\m@ne
  \NextItem=\m@ne
  \MoreFiguresfalse
  \MoreTablesfalse
 \fi
 \ifnum\NextItem=\Figure
   \FindItem{\Figure}{\NextFigure}
   \ifnum\StackPointer=\m@ne \MoreFiguresfalse
   \else
    \GetItemSPAN{\StackPointer}
    \ifnum\ItemSPAN=\Double\relax
     \MoreFiguresfalse\fi
   \fi
   \ifMoreFigures\Print{}\FigureItems\fi
 \fi
 \ifnum\NextItem=\Table
   \FindItem{\Table}{\NextTable}
   \ifnum\StackPointer=\m@ne \MoreTablesfalse
   \else
    \GetItemSPAN{\StackPointer}
    \ifnum\ItemSPAN=\Double\relax
     \MoreTablesfalse\fi
   \fi
   \ifMoreTables\Print{}\TableItems\fi
 \fi
   \MakePageInCompletefalse 
   \ifMoreFigures\MakePageInCompletetrue\fi
   \ifMoreTables\MakePageInCompletetrue\fi
 \repeat
 \ifZoneAFullPage
  \global\TextSize=\z@
  \global\ZoneBSize=\z@
  \global\vsize=\z@\relax
  \global\topskip=\z@\relax
  \vbox to \z@{\vss}
  \eject
 \else
 \global\vsize=\ZoneBSize
 \global\hsize=\ColumnWidth
 \ifdim\TextSize<23pt
 \Warn{}
 \Warn{* Making column fall short: TextSize=\the\TextSize *}
 \vskip-\lastskip\eject\fi
\fi
}

\def\FigureItems{
 \Print{Considering...}
 \ShowItem{\StackPointer}
 \GetItemBOX{\StackPointer} 
 \GetItemSPAN{\StackPointer}
  \CheckFitInZone 
  \ifnum\ItemFits=\Yes
   \ifnum\ItemSPAN=\Single
     \ChangeStatus{\StackPointer}{\InZoneB} 
     \global\FigInZoneBtrue
     \ifFirstSingleItem
      \hbox{}\vskip-\BodgeHeight
     \global\advance\ItemSIZE by \TextLeading
     \fi
     \unvbox\ItemBOX\ItemSep
     \global\FirstSingleItemfalse
     \global\advance\TextSize by -\ItemSIZE
     \global\advance\TextSize by -\TextLeading
   \else
    \ifFirstZoneA
     \global\advance\ItemSIZE by \TextLeading
     \global\FirstZoneAfalse\fi
    \global\advance\TextSize by -\ItemSIZE
    \global\advance\TextSize by -\TextLeading
    \global\advance\ZoneBSize by -\ItemSIZE
    \global\advance\ZoneBSize by -\TextLeading
    \ifFigInZoneB\relax
     \else
     \ifdim\TextSize<3\TextLeading
     \global\ZoneAFullPagetrue
     \fi
    \fi
    \ChangeStatus{\StackPointer}{\Zone}
    \ifnum\Zone=\InZoneC \global\FigInZoneCtrue\fi
  \fi
   \Print{TextSize=\the\TextSize}
   \Print{ZoneBSize=\the\ZoneBSize}
  \global\advance\NextFigure \@ne
   \Print{This figure has been placed.}
  \else
   \Print{No space available for this figure...holding over.}
   \Print{}
   \global\MoreFiguresfalse
  \fi
}

\def\TableItems{
 \Print{Considering...}
 \ShowItem{\StackPointer}
 \GetItemBOX{\StackPointer} 
 \GetItemSPAN{\StackPointer}
  \CheckFitInZone 
  \ifnum\ItemFits=\Yes
   \ifnum\ItemSPAN=\Single
    \ChangeStatus{\StackPointer}{\InZoneB}
     \global\TabInZoneBtrue
     \ifFirstSingleItem
      \hbox{}\vskip-\BodgeHeight
     \global\advance\ItemSIZE by \TextLeading
     \fi
     \unvbox\ItemBOX\ItemSep
     \global\FirstSingleItemfalse
     \global\advance\TextSize by -\ItemSIZE
     \global\advance\TextSize by -\TextLeading
   \else
    \ifFirstZoneA
    \global\advance\ItemSIZE by \TextLeading
    \global\FirstZoneAfalse\fi
    \global\advance\TextSize by -\ItemSIZE
    \global\advance\TextSize by -\TextLeading
    \global\advance\ZoneBSize by -\ItemSIZE
    \global\advance\ZoneBSize by -\TextLeading
    \ifFigInZoneB\relax
     \else
     \ifdim\TextSize<3\TextLeading
     \global\ZoneAFullPagetrue
     \fi
    \fi
    \ChangeStatus{\StackPointer}{\Zone}
    \ifnum\Zone=\InZoneC \global\TabInZoneCtrue\fi
   \fi
  \global\advance\NextTable \@ne
   \Print{This table has been placed.}
  \else
  \Print{No space available for this table...holding over.}
   \Print{}
   \global\MoreTablesfalse
  \fi
}


\def\CheckFitInZone{%
{\advance\TextSize by -\ItemSIZE
 \advance\TextSize by -\TextLeading
 \ifFirstSingleItem
  \advance\TextSize by \TextLeading
 \fi
 \ifnum\Zone=\InZoneA\relax
  \else \advance\TextSize by -\ZoneBAdjust
 \fi
 \ifdim\TextSize<3\TextLeading \global\ItemFits=\No
 \else \global\ItemFits=\Yes\fi}
}

\def\BeginOpening{%
  \ninepoint
  \thispagestyle{titlepage}%
  \global\setbox\ItemBOX=\vbox\bgroup%
    \hsize=\PageWidth%
    \hrule height \z@
    \ifsinglecol\vskip 6pt\fi 
}

\let\begintopmatter=\BeginOpening  

\def\EndOpening{%
  \One
  \egroup
  \ifsinglecol
    \box\ItemBOX%
    \vskip\TextLeading plus 2\TextLeading
    \@noafterindent
  \else
    \ItemNUMBER=\z@%
    \ItemTYPE=\Figure
    \ItemSPAN=\Double
    \ItemSTATUS=\InStack
    \JoinStack
  \fi
}


\newif\if@here  \@herefalse

\def\no@float{\global\@heretrue}
\let\nofloat=\relax 

\def\beginfigure{%
  \@ifstar{\global\@dfloattrue \@bfigure}{\global\@dfloatfalse \@bfigure}%
}

\def\@bfigure#1{%
  \par
  \if@dfloat
    \ItemSPAN=\Double
    \TEMPDIMEN=\PageWidth
  \else
    \ItemSPAN=\Single
    \TEMPDIMEN=\ColumnWidth
  \fi
  \ifsinglecol
    \TEMPDIMEN=\PageWidth
  \else
    \ItemSTATUS=\InStack
    \ItemNUMBER=#1%
    \ItemTYPE=\Figure
  \fi
  \bgroup
    \hsize=\TEMPDIMEN
    \global\setbox\ItemBOX=\vbox\bgroup
      \eightpoint\nostb@ls{10pt}%
      \let\caption=\fig@caption
      \ifsinglecol \let\nofloat=\no@float\fi
}

\def\fig@caption#1{%
  \vskip 5.5pt plus 6pt%
  \bgroup 
    \eightpoint\nostb@ls{10pt}%
    \setbox\TEMPBOX=\hbox{#1}%
    \ifdim\wd\TEMPBOX>\TEMPDIMEN
      \noindent \unhbox\TEMPBOX\par
    \else
      \hbox to \hsize{\hfil\unhbox\TEMPBOX\hfil}%
    \fi
  \egroup
}

\def\endfigure{%
  \par\egroup 
  \egroup
  \ifsinglecol
    \if@here \midinsert\global\@herefalse\else \topinsert\fi
      \unvbox\ItemBOX
    \endinsert
  \else
    \JoinStack
    \Print{Processing source for figure \the\ItemNUMBER}%
  \fi
}


\newbox\tab@cap@box
\def\tab@caption#1{\global\setbox\tab@cap@box=\hbox{#1\par}}

\newtoks\tab@txt@toks
\long\def\tab@txt#1{\global\tab@txt@toks={#1}\global\table@txttrue}

\newif\iftable@txt  \table@txtfalse
\newif\if@dfloat    \@dfloatfalse

\def\begintable{%
  \@ifstar{\global\@dfloattrue \@btable}{\global\@dfloatfalse \@btable}%
}

\def\@btable#1{%
  \par
  \if@dfloat
    \ItemSPAN=\Double
    \TEMPDIMEN=\PageWidth
  \else
    \ItemSPAN=\Single
    \TEMPDIMEN=\ColumnWidth
  \fi
  \ifsinglecol
    \TEMPDIMEN=\PageWidth
  \else
    \ItemSTATUS=\InStack
    \ItemNUMBER=#1%
    \ItemTYPE=\Table
  \fi
  \bgroup
    \eightpoint\nostb@ls{10pt}%
    \global\setbox\ItemBOX=\vbox\bgroup
      \let\caption=\tab@caption
      \let\tabletext=\tab@txt
      \ifsinglecol \let\nofloat=\no@float\fi
}

\def\endtable{%
  \par\egroup 
  \egroup
  \setbox\TEMPBOX=\hbox to \TEMPDIMEN{%
    \eightpoint\nostb@ls{10pt}%
    \hss
    \vbox{%
      \hsize=\wd\ItemBOX
      \ifvoid\tab@cap@box
      \else
        \noindent\unhbox\tab@cap@box
        \vskip 5.5pt plus 6pt%
      \fi
      \box\ItemBOX
      \iftable@txt
        \vskip 10pt%
        \noindent\the\tab@txt@toks
        \global\table@txtfalse
      \fi
    }%
    \hss
  }%
  \ifsinglecol
    \if@here \midinsert\global\@herefalse\else \topinsert\fi
      \box\TEMPBOX
    \endinsert
  \else
    \global\setbox\ItemBOX=\box\TEMPBOX
    \JoinStack
    \Print{Processing source for table \the\ItemNUMBER}%
  \fi
}

\def\UnloadZoneA{%
\FirstZoneAtrue
 \Iteration=\z@
  \loop
   \ifnum\Iteration<\LengthOfStack
    \GetItemSTATUS{\Iteration}
    \ifnum\ItemSTATUS=\InZoneA
     \GetItemBOX{\Iteration}
     \ifFirstZoneA \vbox to \BodgeHeight{\vfil}%
     \FirstZoneAfalse\fi
     \unvbox\ItemBOX\ItemSep
     \LeaveStack{\Iteration}
     \else
     \advance\Iteration \@ne
   \fi
 \repeat
}

\def\UnloadZoneC{%
\Iteration=\z@
  \loop
   \ifnum\Iteration<\LengthOfStack
    \GetItemSTATUS{\Iteration}
    \ifnum\ItemSTATUS=\InZoneC
     \GetItemBOX{\Iteration}
     \ItemSep\unvbox\ItemBOX
     \LeaveStack{\Iteration}
     \else
     \advance\Iteration \@ne
   \fi
 \repeat
}


\def\ShowItem#1{
  {\GetItemAll{#1}
  \Print{\the#1:
  {TYPE=\ifnum\ItemTYPE=\Figure Figure\else Table\fi}
  {NUMBER=\the\ItemNUMBER}
  {SPAN=\ifnum\ItemSPAN=\Single Single\else Double\fi}
  {SIZE=\the\ItemSIZE}}}
}

\def\ShowStack{%
 \Print{}
 \Print{LengthOfStack = \the\LengthOfStack}
 \ifnum\LengthOfStack=\z@ \Print{Stack is empty}\fi
 \Iteration=\z@
 \loop
 \ifnum\Iteration<\LengthOfStack
  \ShowItem{\Iteration}
  \advance\Iteration \@ne
 \repeat
}

\def\B#1#2{%
\hbox{\vrule\kern-0.4pt\vbox to #2{%
\hrule width #1\vfill\hrule}\kern-0.4pt\vrule}
}


\newif\ifsinglecol   \singlecolfalse

\def\onecolumn{%
  \global\output={\singlecoloutput}%
  \global\hsize=\PageWidth
  \global\vsize=\PageHeight
  \global\ColumnWidth=\hsize
  \global\TextLeading=12pt
  \global\Leading=12
  \global\singlecoltrue
  \global\let\onecolumn=\relax
  \global\let\footnote=\sing@footnote
  \global\let\vfootnote=\sing@vfootnote
  \ninepoint 
  \message{(Single column)}%
}

\def\singlecoloutput{%
  \shipout\vbox{\PageHead\vbox to \PageHeight{\pagebody\vss}\PageFoot}%
  \advancepageno
  \ifplate@page
    \shipout\vbox{%
      \sp@pagetrue
      \def\sp@type{plate}%
      \global\plate@pagefalse
      \PageHead\vbox to \PageHeight{\unvbox\plt@box\vfil}\PageFoot%
    }%
    \message{[plate]}%
    \advancepageno
  \fi
  \ifnum\outputpenalty>-\@MM \else\dosupereject\fi%
}

\def\ItemSep{\vskip\ItemSepamount\relax}

\def\ItemSepbreak{\par\ifdim\lastskip<\ItemSepamount
  \removelastskip\penalty-200\ItemSep\fi%
}


\let\@@endinsert=\endinsert 

\def\endinsert{\egroup 
  \if@mid \dimen@\ht\z@ \advance\dimen@\dp\z@ \advance\dimen@12\p@
    \advance\dimen@\pagetotal \advance\dimen@-\pageshrink
    \ifdim\dimen@>\pagegoal\@midfalse\p@gefalse\fi\fi
  \if@mid \ItemSep\box\z@\ItemSepbreak
  \else\insert\topins{\penalty100 
    \splittopskip\z@skip
    \splitmaxdepth\maxdimen \floatingpenalty\z@
    \ifp@ge \dimen@\dp\z@
    \vbox to\vsize{\unvbox\z@\kern-\dimen@}
    \else \box\z@\nobreak\ItemSep\fi}\fi\endgroup%
}


\def\gobbleone#1{}
\def\gobbletwo#1#2{}
\let\footnote=\gobbletwo 
\let\vfootnote=\gobbleone

\def\sing@footnote#1{\let\@sf\empty 
  \ifhmode\edef\@sf{\spacefactor\the\spacefactor}\/\fi
  \hbox{$^{\hbox{\eightpoint #1}}$}\@sf\sing@vfootnote{#1}%
}

\def\sing@vfootnote#1{\insert\footins\bgroup\eightpoint\b@ls{9pt}%
  \interlinepenalty\interfootnotelinepenalty
  \splittopskip\ht\strutbox 
  \splitmaxdepth\dp\strutbox \floatingpenalty\@MM
  \leftskip\z@skip \rightskip\z@skip \spaceskip\z@skip \xspaceskip\z@skip
  \noindent $^{\scriptstyle\hbox{#1}}$\hskip 4pt%
    \footstrut\futurelet\next\fo@t%
}

\def\footnoterule{\kern-3\p@ \hrule height \z@ \kern 3\p@}

\skip\footins=19.5pt plus 12pt minus 1pt
\count\footins=1000
\dimen\footins=\maxdimen

\def\note#1#2{%
  \let\@sf=\empty \ifhmode\edef\@sf{\spacefactor\the\spacefactor}\/\fi
  #1\insert\footins\bgroup
    \eightpoint\b@ls{10pt}\rm
    \interlinepenalty\interfootnotelinepenalty
    \splitmaxdepth\dp\strutbox \floatingpenalty\@MM
    \leftskip\z@skip \rightskip\z@skip \spaceskip\z@skip \xspaceskip\z@skip
    \noindent\footstrut #1$\,$#2\strut\par
  \egroup
  \@sf\relax}


\def\landscape{%
  \global\TEMPDIMEN=\PageWidth
  \global\PageWidth=\PageHeight
  \global\PageHeight=\TEMPDIMEN
  \global\let\landscape=\relax
  \onecolumn
  \message{(landscape)}%
  \raggedbottom
}


\output{%
  \ifLeftCOL
    \global\setbox\LeftBOX=\vbox to \ZoneBSize{\box255\unvbox\ZoneBBOX
      \ifvoid\footins\else
        \vskip\skip\footins\unvbox\footins\fi
    }%
    \global\LeftCOLfalse
    \MakeRightCol
  \else
    \setbox\RightBOX=\vbox to \ZoneBSize{\box255\unvbox\ZoneBBOX
      \ifvoid\footins\else
        \vskip\skip\footins\unvbox\footins\fi
    }%
    \setbox\MidBOX=\hbox{\box\LeftBOX\hskip\ColumnGap\box\RightBOX}%
    \setbox\PageBOX=\vbox to \PageHeight{%
      \UnloadZoneA\box\MidBOX\UnloadZoneC}%
    \shipout\vbox{\PageHead\vbox to \PageHeight{\box\PageBOX\vss}\PageFoot}%
    \advancepageno
    \ifplate@page
      \shipout\vbox{%
        \sp@pagetrue
        \def\sp@type{plate}%
        \global\plate@pagefalse
        \PageHead\vbox to \PageHeight{\unvbox\plt@box\vfil}\PageFoot%
      }%
      \message{[plate]}%
      \advancepageno
    \fi
    \global\LeftCOLtrue
    \CleanStack
    \MakePage
  \fi
}


\Warn{\start@mess}

\newif\ifCUPmtplainloaded 
\ifprod@font
  \global\CUPmtplainloadedtrue
\fi

\def\mnmacrosloaded{} 

\catcode `\@=12 

